\documentclass[10pt,twocolumn,letterpaper]{article}

\usepackage{cvpr}      %

\usepackage{multirow}
\usepackage{graphicx, array, booktabs}
\usepackage{rotating}     %
\usepackage{graphicx,booktabs,multirow,array}

\usepackage{graphicx,array,makecell,subcaption,multirow}

\definecolor{cvprblue}{rgb}{0.21,0.49,0.74}
\usepackage[pagebackref,breaklinks,colorlinks,allcolors=cvprblue]{hyperref}

\def\confName{CVPR}
\def\confYear{2026}

\title{Guided Lensless Polarization Imaging}

\author{
Noa Kraicer \quad Erez Yosef \quad Raja Giryes\\
Tel Aviv University\\
School of Electrical Engineering, Faculty of Engineering\\
{\tt\small noakraicer0@gmail.com \quad  erez.yo@gmail.com \quad raja@tauex.tau.ac.il }
}

\begin{document}
\maketitle

\begin{abstract}
Polarization imaging captures the polarization state of light, revealing information invisible to the human eye yet valuable in domains such as biomedical diagnostics, autonomous driving, and remote sensing. However, conventional polarization cameras are often expensive, bulky, or both, limiting their practical use. Lensless imaging offers a compact, low-cost alternative by replacing the lens with a simple optical element like a diffuser and performing computational reconstruction, but existing lensless polarization systems suffer from limited reconstruction quality.
To overcome these limitations, we introduce a RGB-guided lensless polarization imaging system that combines a compact polarization-RGB sensor with an auxiliary, widely available conventional RGB camera providing structural guidance. We reconstruct multi-angle polarization images for each RGB color channel through a two-stage pipeline: a physics-based inversion recovers an initial polarization image, followed by a Transformer-based fusion network that refines this reconstruction using the RGB guidance image from the conventional RGB camera.
Our two-stage method significantly improves reconstruction quality and fidelity over lensless-only baselines, generalizes across datasets and imaging conditions, and achieves high-quality real-world results on our physical prototype lensless camera without any fine-tuning.
\end{abstract}

\begin{figure}[t]
    \centering
    \setlength{\tabcolsep}{1.5pt} %
    \renewcommand{\arraystretch}{0}

    \begin{tabular}{cc}
        \begin{subfigure}[t]{0.36\linewidth}
            \centering
            \includegraphics[width=\linewidth,height=\linewidth]{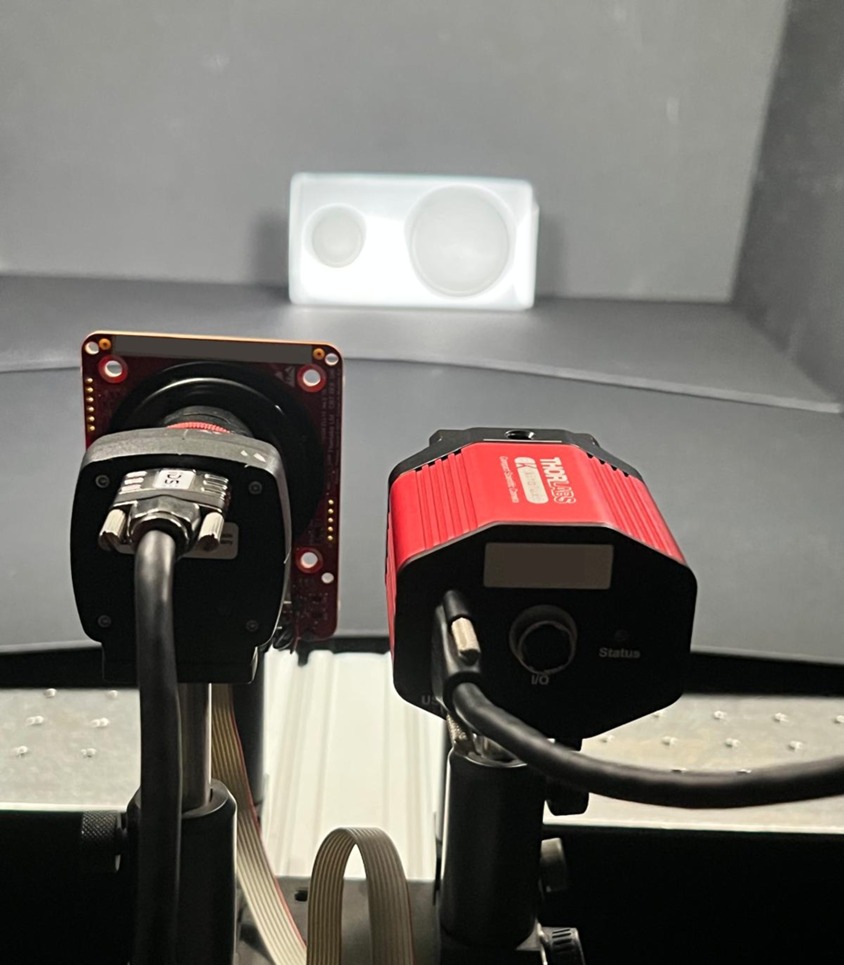}
            \caption{Our optical setup}
        \end{subfigure} &
        \begin{subfigure}[t]{0.36\linewidth}
            \centering
            \includegraphics[width=\linewidth,height=\linewidth]{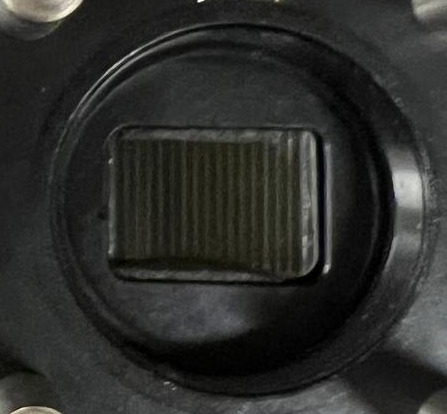}
            \caption{Self-fabricated polarization mask}
        \end{subfigure} \\
        \begin{subfigure}[t]{0.36\linewidth}
            \centering
            \includegraphics[width=\linewidth,height=\linewidth]{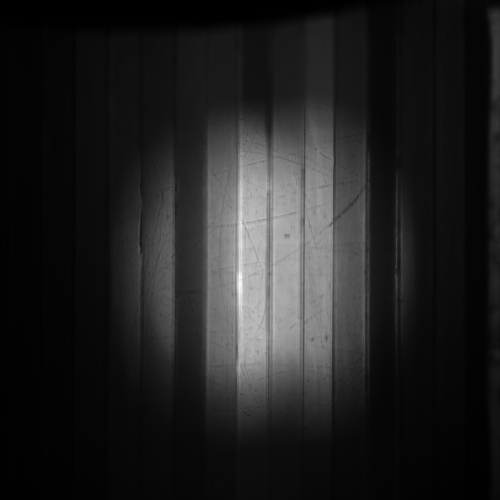}
            \caption{Captured lensless polarization image}
        \end{subfigure} &
        \begin{subfigure}[t]{0.36\linewidth}
            \centering
            \includegraphics[width=\linewidth,height=\linewidth]{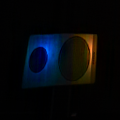}
            \caption{Reconstructed polarization result}
        \end{subfigure}
    \end{tabular}

    \vspace{-5pt}
    \caption{\textbf{RGB-guided lensless polarization imaging system:} (a) optical setup; (b) custom polarization mask; (c) captured lensless image under front illumination with two orthogonally polarized projectors; and (d) reconstructed grayscale polarization result, visualized by mapping the \(0^\circ\), \(45^\circ\), and \(90^\circ\) outputs to the R, G, and B channels.}
    \label{fig:lensless_reco_setup}
\end{figure}

\section{Introduction}
\label{sec:intro}
Polarization is a fundamental property of light that describes the orientation and phase relationship between its orthogonal electric-field components. Because reflections and material anisotropy affect the polarization state, it encodes information about surface geometry, reflectance, and composition, details often inaccessible to standard intensity or RGB imaging \cite{tyo2006review,huynh2010shape}.
\begin{figure*}[t]
\centering
\includegraphics[width=0.99\textwidth]{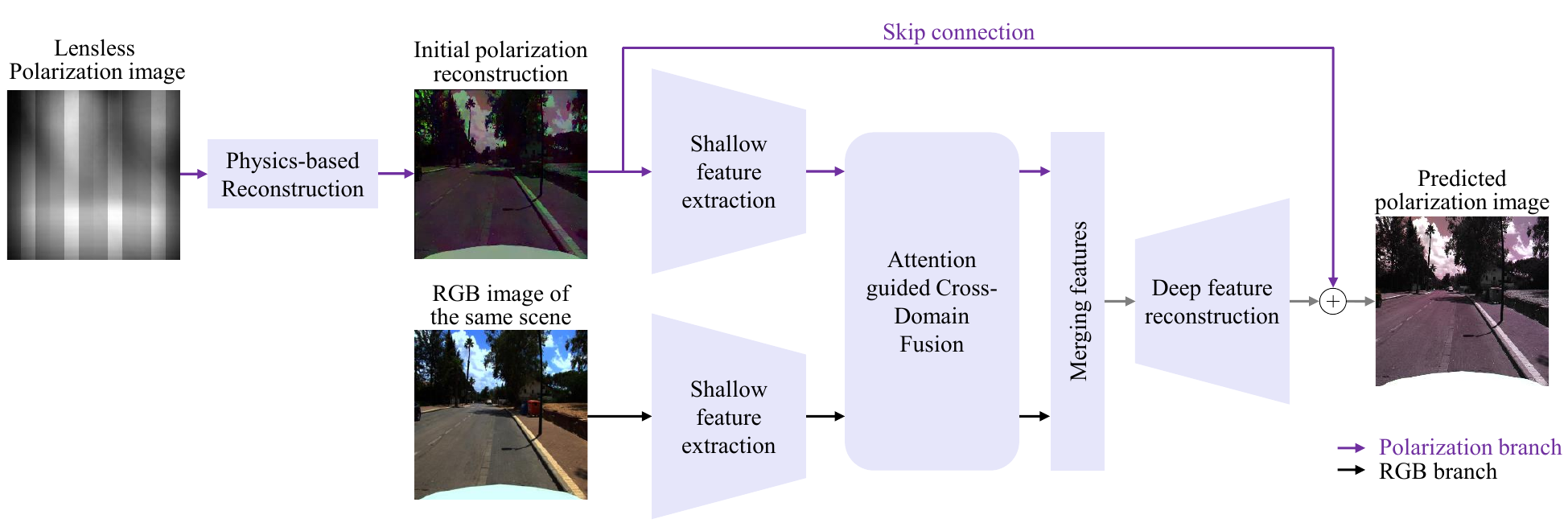} 
\vspace{-0.1in}
\caption{Overview of the proposed RGB-guided reconstruction pipeline. The process consists of two stages: (1) polarization intensity images (color or grayscale) are reconstructed from lensless measurements using a physics-based algorithm (FISTA/ADMM); and (2) the initial reconstruction and a registered RGB image of the same scene are separately encoded and fused through cross-domain attention to produce a refined polarization reconstruction. For visualization, the grayscale reconstructions at three polarization angles ($0^\circ$, $45^\circ$, $90^\circ$) are mapped to the R, G, and B channels. The pipeline is compatible with more general input configurations.}
\vspace{-0.1in}

\label{fig:proposed_pipeline}
\end{figure*}
Recent approaches to polarization imaging include division-of-focal-plane (DoFP) sensors, which assign different polarization filters to adjacent pixels; time-sequential polarimetry, which uses rotating or tunable elements to capture multiple polarization states over time \cite{tyo2006review}, and lensless polarization imaging, which replaces conventional lenses with spatially coded optical elements and reconstructs the image computationally.
Despite its utility, polarization imaging has yet to achieve widespread adoption, largely due to the cost, size, and complexity of conventional polarization cameras.

Lensless imaging offers a compelling alternative by replacing lenses with simple optical elements, such as diffusers or coded masks, and shifting hardware complexity to computation. Such schemes enable compact, low-cost, and scalable imaging systems \cite{boominathan2021recent}. 
Recent works have extended lensless imaging to polarization by combining polarization-sensitive components with various optical coding schemes \cite{kraicer2026lensless,elmalem2021lensless,baek2022lensless,wang2025lensless}. Yet, reconstruction quality remains limited due to the highly compressed measurements that jointly encode structural and polarization information, compounded by the inherently ill-posed nature of the inverse problem.

Classical model-based solvers, such as the Fast Iterative Shrinkage-Thresholding Algorithm (FISTA) \cite{beck2009fast} and the Alternating Direction Method of Multipliers
(ADMM) algorithm \cite{boyd2011distributed}, can recover coarse polarization estimates but often fail to recover high-frequency details and are sensitive to noise and deviations from the assumed imaging model. While deep learning models have significantly improved reconstruction quality in general lensless imaging tasks \cite{boominathan2021recent}, to the best of our knowledge, they have not yet been explored for lensless polarization imaging.

RGB images provide complementary geometric and edge information consistent across polarization states, offering a natural way to regularize the ill-posed inversion and recover fine details. While adding an RGB camera increases system complexity compared to a fully lensless design, RGB cameras are compact, low-cost, and widely available, keeping the system simpler and more affordable than dedicated polarization cameras. This is particularly important for size- and cost-constrained applications such as compact microscopy and endoscopy \cite{polMicroscopy,polImg_medical,polTissueRec,WallerMiniScope}.

Leveraging the complementary geometric information provided by RGB images, we introduce the \textbf{first RGB-guided lensless polarization imaging system} that integrates physics-based reconstruction with data-driven refinement (\Cref{fig:proposed_pipeline}), enabling accurate recovery of fine structures and details.
The first stage performs a model-based inversion (e.g., FISTA or ADMM) to obtain a physically consistent initialization from raw lensless measurements.
The second stage refines this estimate using a Swin Transformer–based fusion network, based on SwinFuSR \cite{arnold2024swinfusr}, which fuses polarization features from the initial reconstruction with RGB features from a lensed camera via alternating self- and cross-attention.
Our approach generalizes across diverse scenes and imaging conditions, and achieves strong performance on real-world measurements without additional fine-tuning.

Our main contributions may be summarized as follows:
(i) We propose the first \textbf{RGB-guided} lensless polarization imaging system, combining simple and low-cost hardware with a reconstruction algorithm achieving state-of-the-art results for lensless polarization imaging;
(ii) We design a two-stage reconstruction approach that integrates a physics-based solver (e.g., FISTA/ADMM) with an adapted version of Swin Transformer utilized for cross-modal fusion, enabling RGB-guided reconstruction of polarization intensity images through self- and cross-attention;
(iii) We conduct extensive experiments on multiple simulated datasets, demonstrating consistent improvements over lensless-only baselines in PSNR, SSIM, and LPIPS, with strong generalization across datasets and unseen point-spread-functions (PSFs);
(iv) We demonstrate promising real-world results on a prototype lensless polarization camera, validating the method’s practical feasibility without additional fine-tuning.

\noindent\textbf{Project page with code, pretrained models, and data:} \url{https://noa-kraicer.github.io/Guided-Lensless-Polarization-Imaging/}.
\section{Related Work}
\label{sec:relatework}
\noindent \textbf{Polarization Imaging.}
Most existing polarization cameras are based on sequential filtering or division-of-focal-plane (DoFP) architectures \cite{tyo2006review}. DoFP sensors enable single-shot capture but suffer spatial resolution loss due to pixel subdivision. Sequential filtering preserves full resolution via multiple exposures with rotating or switching polarizers but introduces motion artifacts and mechanical complexity, limiting real-time use.
Recent designs address these trade-offs using stacked polarizer architectures \cite{sasagawa2022polarization} and flat-optics or metasurface-based polarization elements \cite{huang2023high, zuo2023chip, li2025flat}, signaling a shift toward miniaturized polarization cameras and motivating exploration of lensless variants.

\noindent \textbf{Lensless Imaging.}
Lensless imaging replaces traditional lenses with coded optics and computational reconstruction, enabling compact imaging systems. Several designs have been proposed, including FlatCam \cite{asif2016flatcam}, which uses a static amplitude mask; DiffuserCam \cite{antipa2018diffusercam}, based on a diffuser and compressive sensing; phase masks \cite{boominathan2020phlatcam}; and programmable optics \cite{zomet2006lensless, miller2020particle,hua2020sweepcam,zheng2021simple,huang2013lensless}. A comprehensive review of lensless imaging systems can be found in \citet{boominathan2022recent}.
Subsequent research has extended lensless imaging to additional modalities, including hyperspectral \cite{monakhova2020spectral}, depth \cite{bagadthey2022flatnet3d}, and temporal imaging \cite{antipa2019video}, highlighting the versatility of the approach. However, lensless polarization imaging has received limited attention and introduces a more complex forward model.

\noindent \textbf{Lensless Polarization Imaging}
Combining lensless imaging with polarization sensing enables compact, multi-modal imaging systems. Prior works have explored various hardware configurations for single-shot capture, including the use of diffusers with polarization mask~\cite{kraicer2026lensless,elmalem2021lensless}, phase masks and polarization-encoded apertures~\cite{baek2022lensless}, and coded masks paired with DoFP polarization sensor~\cite {wang2025lensless}, 

Despite these hardware advances, accurate image reconstruction remains a major challenge. The highly compressed and multiplexed nature of the measurements leads to severely ill-posed inverse problems that are sensitive to noise, artifacts, and other real-world imperfections.

\noindent \textbf{Lensless Imaging Reconstruction.}
Classical lensless reconstruction methods formulate image recovery as a variational inverse problem solved by iterative optimization algorithms such as FISTA, ADMM, and their variants~\cite{antipa2018diffusercam, monakhova2020spectral, boominathan2020phlatcam}.  
For lensless polarization imaging, \citet{elmalem2021lensless} use TV-regularized FISTA, while \citet{baek2022lensless} and \citet{wang2025lensless}
 adopt ADMM-based solvers.
These physics-driven methods are interpretable but remain sensitive to noise and PSF mismatch, and cannot fully recover fine details.
Deep learning has advanced lensless imaging reconstruction from classical physics-based optimization to learned data-driven models. Early approaches learn direct mappings from sensor measurements to images using convolutional neural networks~\cite{sinha2017lensless}, while later methods integrate model-based priors through unrolled or hybrid architectures such as ISTA-Net~\cite{zhang2018ista}, ADMM-Net~\cite{sun2016deep}. Hybrid systems like FlatNet \cite{khan2019towardsflatnet,khan2020flatnet}, DifuzCam \cite{yosef2025difuzcam}, and GANESH \cite{madavan2025ganesh} combine physical inversion with learned refinement for improved fidelity and generalization.  
Recent modular frameworks further enhance robustness under domain shifts \cite{bezzam2025towards} or employing efficient adaptations for new distributions \cite{yismaw2024domain}.

However, existing approaches for lensless \emph{polarization} imaging remain purely physics-based, relying on explicit forward models, iterative optimization, and handcrafted priors. Although learning based approaches have shown promise for polarization demosaicing \cite{pistellato2022deep,zeng2019end}, denoising \cite{li2020learning}, deblurring \cite{zhou2025learning}, and low-light enhancement \cite{hu2020iplnet, xu2022colorpolarnet}, these efforts focus on lens-based systems and have not been demonstrated for lensless polarization reconstruction.

\noindent \textbf{Cross-Modal Guidance.}
Cross-modal fusion has proven to be effective in enhancing degraded modalities using complementary ones. 
SwinFuSR \cite{arnold2024swinfusr} uses high-resolution RGB images to guide thermal super-resolution via a Swin Transformer, while \citet{yosef2025tell} presented an RGB image denoising using scene textual description with a diffusion model.
In polarization imaging, \citet{liu2023polarization} uses RGB guidance to demosaic simulated sparse DoFP data and recover Stokes parameters.
PolarFree \cite{yao2025polarfree} shows that polarization can act as a powerful auxiliary signal for reflection removal in RGB images. PolarAnything \cite{zhang2025polaranything} generates polarization images directly from RGB inputs using diffusion models; however, it relies solely on learned priors and lacks physical fidelity. Yet, RGB guided reconstruction for \emph{lensless polarization} imaging has not been explored.
We address this gap with a two-stage framework that leverages RGB guidance to enhance reconstruction fidelity and robustness.

\section{Method}
\label{sec:method}
Our framework reconstructs high-quality polarization intensity images from a single-shot lensless measurement, leveraging guidance from a registered RGB image. We first formulate the lensless polarization imaging setup (\ref{subsec:problem}), and describe the synthetic data generation process used for the training (\ref{subsec:simulation}). We then describe our two-stage reconstruction pipeline: a physics-based optimization stage (\ref{subsec:fista}) and an RGB-guided transformer-based refinement stage (\ref{subsec:network}). \Cref{fig:proposed_pipeline} provides an overview of the full pipeline.

\begin{figure}[t]
\centering
\includegraphics[width=\linewidth]{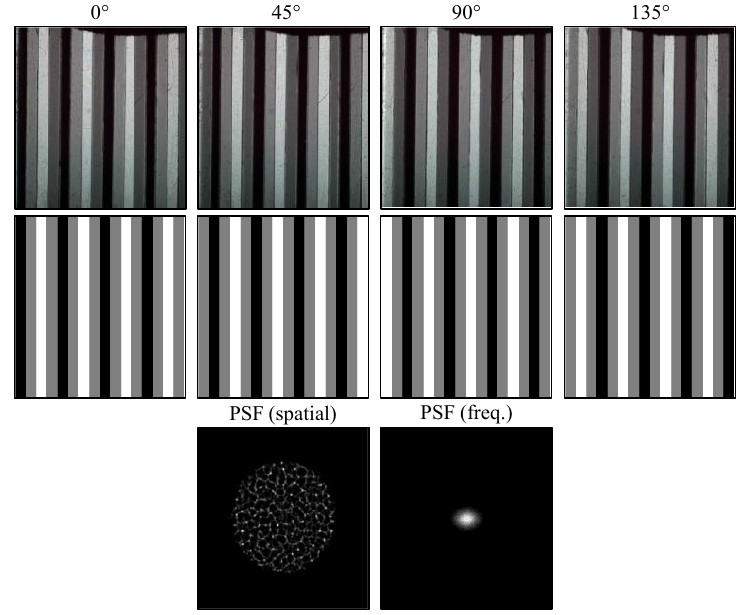}
\vspace{-0.15in}
\caption{Real and simulated polarization mask responses for four polarization angles in grayscale. (Top) Real masks captured with our prototype; (center) simulated masks; and (bottom) measured grayscale PSF of our diffuser in spatial and frequency domains. The PSF is low-pass, leading to loss of high-frequency information.}
\label{fig:polmask}
\end{figure}

\subsection{Lensless Polarization Imaging Setup}
\label{subsec:problem}
We employ a lensless imaging system that captures a spatio-polarimetrically multiplexed measurement in a single shot. Our optical design, adapted from \cite{kraicer2026lensless,elmalem2021lensless}, combines a diffuser PSF with a striped polarization mask composed of linear polarizers oriented at $0^\circ$, $45^\circ$, $90^\circ$, and $135^\circ$ on an RGB sensor. Placed directly on the sensor, this mask spatially encodes the polarization state of incident light by transmitting different polarization components through distinct stripe orientations according to Malus' law \cite{brukner1999malus}, as illustrated in \Cref{fig:polmask}. The diffuser mixes this information in the captured image, but due to its frequency response, it loses the high-frequency information.
A detailed description of the optical setup is provided in the Supplementary Material.
The resulting measurement can be modeled as a linear inverse problem: $\mathbf{y} = A \mathbf{x} + \mathbf{e}$,
where $\mathbf{y}$ denotes the captured lensless measurement, $\mathbf{x}$ is the multi-angle polarization intensity image to be reconstructed, and $A$ is the forward imaging operator combining the effects of the diffuser and polarization mask. The additive term $\mathbf{e}$ accounts for measurement noise.
Due to the strong spatial multiplexing, loss of high frequencies, and partial polarization sampling, this inverse problem is highly ill-posed. To address this, we use an additional conventional RGB camera capturing the same scene, and leverage the RGB image as guidance to provide structural cues and recover high-frequency details.

\begin{figure*}[t]
\centering
\setlength{\tabcolsep}{2pt}   %
\renewcommand{\arraystretch}{1.15} %

\begin{tabular}{>{\centering\arraybackslash}m{0.04\textwidth}
                *{6}{>{\centering\arraybackslash}m{0.145\textwidth}}}

& {\normalsize {RGB (guide)}} &
  {\normalsize {FISTA pred}} &
  {\normalsize {FISTA + Transformer}} &
  {\normalsize {Ours}} &
  {\normalsize {Ours (fine-tuned)}} &
  {\normalsize {GT}} \\[2pt]

\multirow{2}{*}{\raisebox{-0.8
in}{\rotatebox{90}{\normalsize UPLight}}} & 
\includegraphics[width=\linewidth]{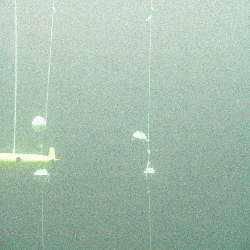} &
\includegraphics[width=\linewidth]{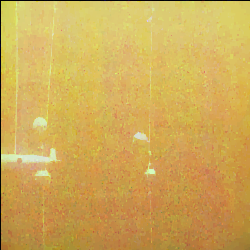} &
\includegraphics[width=\linewidth]{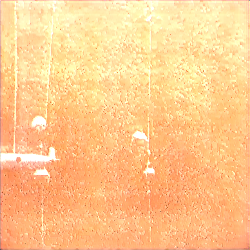} &
\includegraphics[width=\linewidth]{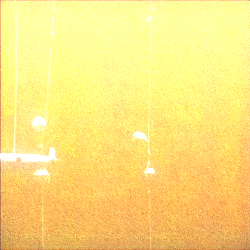} &
\includegraphics[width=\linewidth]{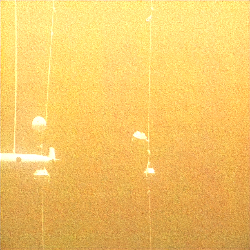} &
\includegraphics[width=\linewidth]{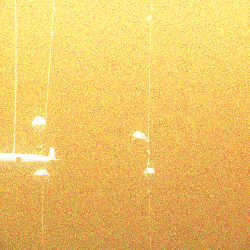} \\[5pt]

& %
\includegraphics[width=\linewidth]{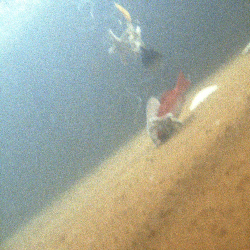} &
\includegraphics[width=\linewidth]{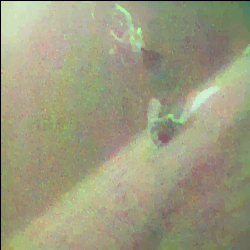} &
\includegraphics[width=\linewidth]{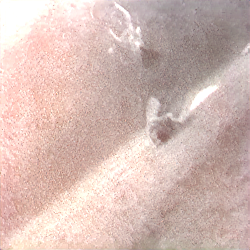} &
\includegraphics[width=\linewidth]{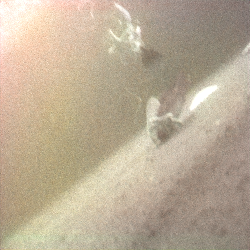} &
\includegraphics[width=\linewidth]{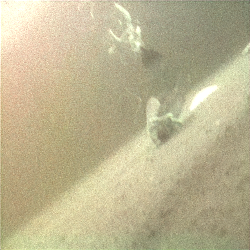} &
\includegraphics[width=\linewidth]{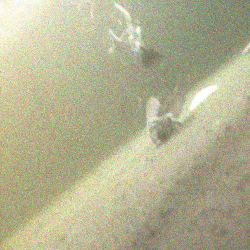} \\[10pt]

\multirow{2}{*}{\raisebox{-0.8
in}{\rotatebox{90}{\normalsize ZJU-RGB-P}}} & 

\includegraphics[width=\linewidth]{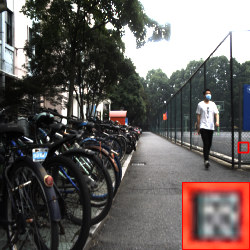} &
\includegraphics[width=\linewidth]{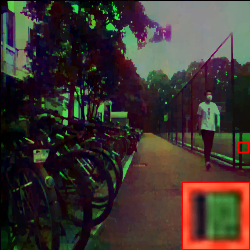} &
\includegraphics[width=\linewidth]{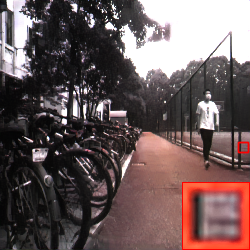} &
\includegraphics[width=\linewidth]{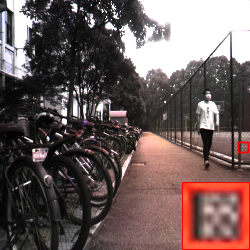} &
\includegraphics[width=\linewidth]{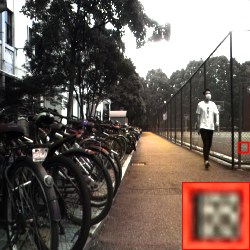} &
\includegraphics[width=\linewidth]{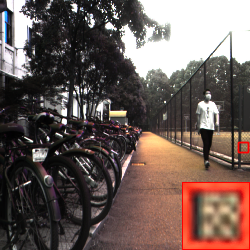} \\[5pt]

&
\includegraphics[width=\linewidth]{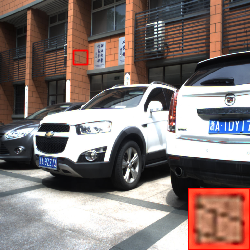} &
\includegraphics[width=\linewidth]{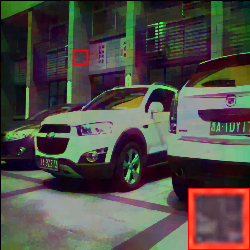} &
\includegraphics[width=\linewidth]{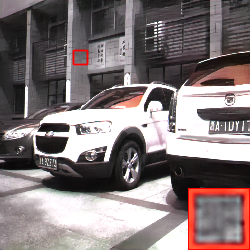} &
\includegraphics[width=\linewidth]{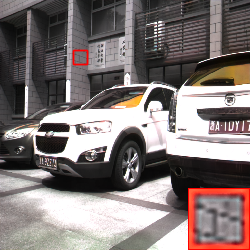} &
\includegraphics[width=\linewidth]{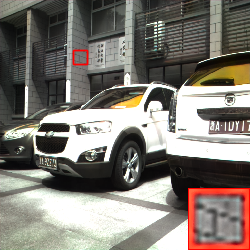} &
\includegraphics[width=\linewidth]{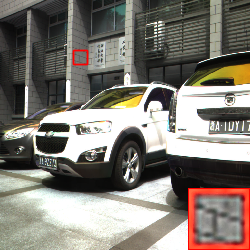} 
\end{tabular}
\vspace{-0.1in}
\caption{Qualitative reconstruction results on \textbf{UPLight} and \textbf{ZJU-RGB-P}. Columns: RGB guidance, FISTA reconstruction, FISTA + Transformer w/o RGB, our full RGB-guided model, its fine-tuned version, and the ground-truth polarization image. Each  polarization grayscale triplet (\(0^\circ\), \(45^\circ\), \(90^\circ\)) is visualized as an RGB composite. Note how the RGB guidance improves the high-frequency recovery.} 
\label{fig:sim_res}
\vspace{-2mm}
\end{figure*}

\noindent \textbf{Forward Model.}
Our lensless polarization camera follows the same compressive imaging principles as the Spectral DiffuserCam \cite{monakhova2020spectral}.
The polarization-independent diffuser convolves the scene with its PSF, where each narrow feature acts like a micro-lens, mapping a point source to a point on the sensor and spatially multiplexing light from all scene points (\Cref{fig:polmask}). This multiplexing enables reconstruction from a subset of sensor pixels, allowing the polarization mask to perform partial sampling across polarization angles. The mask transmits light according to its orientation, applying a multiplicative modulation to the incident intensity. While the diffuser’s PSF enables recovery with the mask, it loses some high-frequency information.
Let $\mathbf{x} \in \mathbb{R}^{H \times W \times C \times P}$ denote the intensity of
the scene per angle of polarization, where $H$ and $W$ are the spatial dimensions, $C \in \{1,3\}$ is the number of
color channels (grayscale or RGB), and $P \in \{3,4\}$ corresponds to angles $0^\circ,45^\circ,90^\circ$ (and optionally $135^\circ$). Each $\mathbf{x}_{:,:,c,p}$ is the intensity that would be observed after an ideal polarizer at angle $p$.
Accordingly, the polarization mask is modeled as a binary spatial selector $\mathbf{S}_p$ that assigns sensor regions to each orientation, since the angular dependence is already encoded in $\mathbf{x}_{:,:,c,p}$.
For each color channel $c$, the measurement is:
\begin{equation}
\label{eq:forward_model}
\mathbf{y}_{:,:,c}
= \sum_{p=1}^{P}
\mathbf{S}_{p}\,\odot\,
\bigl(\mathbf{x}_{:,:,c,p} * \mathbf{k}_c\bigr)
\end{equation}
where $\mathbf{k}_c$ is the diffuser PSF for channel $c$, $\odot$ denotes element-wise
multiplication, and $*$ denotes 2D convolution over spatial dimensions.

\subsection{Synthetic Data Generation}
\label{subsec:simulation}
Training deep neural networks requires a large number of labeled samples, which is impractical to collect from a real lensless polarization camera at scale. Moreover, to the best of our knowledge, there are no publicly available datasets specifically designed for lensless polarization reconstruction. To address this, we simulate lensless measurements from existing datasets: Polarimetric Imaging for Perception (PIP)~\cite{baltaxe2023polarimetric}, UPLight~\cite{liu2025sharecmp}, and ZJU-RGB-P~\cite{xiang2021polarization}. Each dataset provides pixel-level aligned RGB-polarimetric data with four polarization orientations per scene, from which we generate synthetic lensless measurements using the forward model in \Cref{eq:forward_model}.

An equivalent unpolarized RGB image is computed as:
\begin{equation}
\mathbf{f}_{\mathrm{RGB}} = \frac{1}{2}(\mathbf{I}_{0^\circ} + \mathbf{I}_{45^\circ} + \mathbf{I}_{90^\circ} + \mathbf{I}_{135^\circ}),
\end{equation}
representing the total unpolarized intensity~\cite{liu2023polarization}.
For simulation, we adopt two configurations:
\begin{enumerate}
    \item a three-angle grayscale setup capturing dominant polarization behavior~\cite{wolff2002constraining}.
    \item a four-angle RGB configuration matching our hardware and improving robustness and noise reduction~\cite{lefaudeux2008compact}.
\end{enumerate} The simulated polarization mask reproduces the prototype’s periodic structure while omitting fabrication artifacts (e.g., dust, edge roughness), enabling consistent data generation independent of a specific physical mask's imperfections.
Each mask pattern consists of four repeated vertical stripe sequences (manufacturing–quality tradeoff). It is applied identically across color channels to maintain consistency with the hardware design.
For realism, we use the PSF measured from our physical device for the simulation process.
\Cref{fig:polmask} illustrates the four-angle configuration, shown in grayscale for visualization, along with the corresponding measured PSF and hardware mask for direct comparison.

\setlength{\tabcolsep}{1.5mm}
\begin{table*}[t]
\centering
\small
\caption{Quantitative results (PSNR $\uparrow$ / SSIM $\uparrow$ / LPIPS $\downarrow$) 
for both \textbf{four-angle RGB} and \textbf{three-angle grayscale} configurations.}
\label{tab:results}
\begin{tabular}{llccccccccc}
\hline
\textbf{Modality} & \textbf{Model} & 
\multicolumn{3}{c}{\textbf{PIP}} & 
\multicolumn{3}{c}{\textbf{UPLight}} &
\multicolumn{3}{c}{\textbf{ZJU-RGB-P}} \\
 &  & PSNR$\uparrow$ & SSIM$\uparrow$ & LPIPS$\downarrow$ & PSNR$\uparrow$ & SSIM$\uparrow$ & LPIPS$\downarrow$ & PSNR$\uparrow$ & SSIM$\uparrow$ & LPIPS$\downarrow$ \\
\hline
\multirow{6}{*}{Color} 
& FISTA               & 14.76 & 0.47 & 0.44 & 12.20 & 0.18 & 0.60 & 15.27 & 0.48 & 0.45 \\
& ADMM & 12.92 & 0.30 & 0.57 & 10.93 & 0.14 & 0.65 & 14.69 & 0.34 & 0.53 \\
& FISTA + Transf.      & 28.31 & 0.86 & 0.12 & 15.98 & 0.37 & 0.36 & 25.65 & 0.86 & 0.15 \\
& ADMM + Transf. & 25.41  & 0.81 & 0.17 & 14.63 & 0.35 & 0.41 & 22.38 & 0.79 & 0.22 \\
& Ours (ADMM input) & 32.48 & \textbf {0.95} & \textbf {0.04} & 19.63 & \textbf {0.51} & 0.29 &  29.11 & \textbf {0.96} & \textbf {0.04} \\
& Ours (FISTA input) & \textbf{33.05} & \textbf{0.95} & \textbf{0.04} & \textbf{20.06} & \textbf{0.51} & \textbf {0.28} & \textbf{30.38} & \textbf{0.96} & \textbf{0.04} \\
\hline
\multirow{6}{*}{Grayscale} 
& FISTA               & 13.87 & 0.45 & 0.45 & 16.72 & 0.26 & 0.53 & 14.50 & 0.46 & 0.44 \\
& ADMM & 13.06 & 0.31 & 0.63 & 11.98 & 0.18 & 0.71 & 14.90 & 0.36 & 0.59 \\
& FISTA + Transf.      & 28.85 & 0.88 & 0.12 & 17.93 & 0.44 & 0.53 & 27.20 & 0.89 & 0.19 \\
& ADMM + Transf. & 24.87  & 0.81 & 0.20 & 15.37 & 0.40 & 0.76 & 23.32 & 0.80 & 0.29 \\
& Ours (ADMM input) & 34.40 & \textbf {0.97}&\textbf {0.03} & 18.45 & \textbf {0.53} & 0.46 &  {29.44} &\textbf {0.97} & 0.09 \\
& Ours (FISTA input)& \textbf{35.13} & \textbf{0.97} & \textbf{0.03} & \textbf{20.49} & 0.52 & \textbf{0.32} & \textbf{31.19} & \textbf{0.97} & \textbf{0.07} \\
\hline
\end{tabular}
\end{table*}

\subsection{Stage I: Physics-based reconstruction}
\label{subsec:fista}
Recovering the polarization intensity image $\hat{\mathbf{x}}$ is achieved by solving the optimization problem:
\begin{equation}
\hat{\mathbf{x}} = \arg\min_{\mathbf{x}} \frac{1}{2\sigma_e^2} \left\| \mathbf{y} - A \mathbf{x} \right\|_2^2 + s(\mathbf{x}),
\end{equation}
where $A$ denotes the forward operator (\Cref{eq:forward_model}), and $\sigma_e$ is the standard deviation of measurement noise. 
The first term enforces fidelity to the sensor measurement, and \( s(\mathbf{x}) \) is a regularization term promoting desirable image priors. The unknown polarization image is represented as \( \mathbf{x} \in \mathbb{R}^{H \times W \times C \times P} \), while the observed measurement \( \mathbf{y} \in \mathbb{R}^{H \times W \times C} \) corresponds to either a real sensor acquisition or a simulated observation generated as described in \Cref{subsec:simulation}.
We solve this inverse problem using FISTA~\cite{beck2009fast} with a fixed number of iterations, employing a weighted 3D Total Variation (3DTV) prior~\cite{kamilov2016parallel} to enforce smoothness across spatial and polarization dimensions, along with a non-negativity constraint.
For simulated data, we use the same PSF and polarization mask as in the simulation, while for real measurements, we use the device-measured PSF and mask shown in \Cref{fig:polmask}. We also implement an alternative solver based on ADMM~\cite{boyd2011distributed} using the same prior, demonstrating the generality of our formulation. 
Additional implementation details, including iteration settings, are provided in the supplementary.
  
\subsection{Stage II: RGB-Guided Deep Refinement}
\label{subsec:network}
While stage I provides a coarse polarization estimate $\hat{\mathbf{x}}$, it lacks fine spatial details due to the loss of high frequencies (see \Cref{fig:polmask}). To recover the high-frequency details and improve reconstruction quality, we employ an RGB-guided refinement network based on SwinFuSR~\cite{arnold2024swinfusr}, a dual-branch Transformer originally designed for RGB-guided thermal super-resolution.  

We adapt SwinFuSR to our task by (i) modifying input/output channels for our polarization data; (ii)  training on full-resolution images instead of patches to correct global, spatially correlated artifacts from FISTA/ADMM reconstruction;  and (iii) incorporating an LPIPS perceptual loss \cite{zhang2018unreasonable} to enhance perceptual quality, consistently with prior SR and reconstruction studies \cite{cai2024phocolens, zeng2021robust,ma2020structure}. 
To improve robustness to small registration errors in real-world measurements, we apply random translation augmentation during training, synthetically shifting both the RGB image and the ground-truth polarization image jointly by up to $\pm 4$ pixels in both horizontal and vertical directions.

The network processes $\hat{\mathbf{x}}$ and the approximately aligned (with residual shifts of up to $\pm 4$ pixels) RGB image $\mathbf{f}_{\mathrm{RGB}}$ through separate branches of shallow convolutional layers and Swin Transformer layers (STL). Feature fusion is later performed using Attention-guided Cross-domain Fusion (ACF) blocks, which alternate between self-attention and cross-attention to integrate information from both modalities. The outputs of the two branches are then merged via concatenation followed by convolution. 
The fused features are subsequently refined through additional STLs and convolutional layers, where a skip connection adds the initial reconstruction to the output to preserve consistency with our physical forward model. Further implementation details appear in the supplementary.

\section{Experimental Results}
\label{sec:results}
We evaluate our proposed method on both synthetic and real-world data. For quantitative evaluation, we used PSNR, 
SSIM~\cite{wang2004image} and LPIPS~\cite{zhang2018unreasonable}. Our RGB-guided method outperforms existing state-of-the-art methods and an equivalent non-guided approach across all tested datasets and metrics.
We demonstrate the advantages of our method on both synthetic datasets and real-world data using our prototype camera. We also present ablation studies to analyze the impact of PSF variations, initialization steps, and fusion strategies.

\noindent \textbf{Implementation Details.}
We train our model on the PIP dataset~\cite{baltaxe2023polarimetric}, which is the largest RGB–polarization dataset ($\sim$12.6K samples). We split the data into training, validation, and test sets with no scene overlap.
Training is conducted under the two configurations as earlier described in the simulation process (\Cref{subsec:simulation}) with aligned RGB guidance images. We use AdamW~\cite{loshchilov2017decoupled} with a OneCycle schedule~\cite{smith2019super} (peak $1.5\times10^{-4}$) for 30 epochs.
The loss combines an $\ell_1$ term and LPIPS~\cite{zhang2018unreasonable}, with weights of 1.0 and 0.1, respectively, with early stopping on validation loss. Inputs for training and evaluation are resized to $250\times250$ pixels and run on an NVIDIA RTX 2080~Ti.
Translation augmentation is used only during training of the RGB-guided models. All reported results are obtained without augmentation.
\subsection{Synthetic Data Results}
\label{sec:results:synthetic}
We compare our approach against two physics-based reconstruction baselines: 
3D FISTA with total variation (TV) regularization~\cite{monakhova2020spectral}, and 3D ADMM variant with similar regularization. Both are widely used in lensless polarization reconstruction. 
In addition, we evaluate a learning-based baseline derived from our architecture, denoted \textit{FISTA/ADMM + Transformer}, in which the refinement network operates solely on the initial reconstruction, without RGB guidance. This is implemented by feeding the same input to both branches and removing cross-modal fusion, thereby disentangling the contribution of RGB guidance.
We also compare against two further learning-based baselines: FlatNet~\cite{khan2020flatnet}, which employs a learnable inversion followed by a perceptual refinement U-net, and PolarAnything~\cite{zhang2025polaranything} that synthesizes polarimetric observations from RGB inputs based on Stable Diffusion v1.5 ~\cite{rombach2022high}.
We adapt PolarAnything to predict polarization \emph{intensity} images instead of the original angle and degree of linear polarization (AoLP/DoLP), testing two conditioning modes— an RGB image and the FISTA reconstruction. Both FlatNet and PolarAnything are evaluated under the \textbf{three-angle grayscale} configuration, since their architectures do not natively support multi-channel RGB–polarization inputs, and are retrained on our dataset for fairness.
\Cref{tab:different_baselines} shows that neither of them performs well on the PIP test set. FlatNet’s learned deconvolution struggles under partial polarization sampling, lacking the iterative regularization and measurement consistency of physics-based solvers. PolarAnything, despite its strong generative prior, performs poorly when adapted to polarization intensity prediction and conditioned on either RGB or FISTA inputs, failing to match the accuracy of our method with or without RGB guidance.
They both fail to generalize to the unseen datasets (see Tab.~1, supplementary).

\begin{table}[t]
\centering
\caption{Comparison of FlatNet, PolarAnything, FISTA + Transformer, and our method under the three-angle grayscale configuration on the PIP dataset.}
\label{tab:different_baselines}
\begin{tabular}{lccc}
\toprule
\textbf{Model} & PSNR$\uparrow$ & SSIM$\uparrow$ & LPIPS$\downarrow$ \\
\midrule
FlatNet & 21.57 & 0.68 & 0.45 \\
PolarAnything (FISTA input) & 21.51 & 0.64 & 0.31 \\
PolarAnything (RGB input) & 22.02 & 0.66 & 0.29 \\
FISTA + Transf. & 28.85 & 0.88 & 0.12 \\
Ours (FISTA input) & \textbf{35.53} & \textbf{0.97} & \textbf{0.03} \\
\bottomrule
\end{tabular}
\end{table}

The remaining methods (excluding FlatNet and PolarAnything) are evaluated on two additional datasets: ZJU-RGB-P~\cite{xiang2021polarization} (394 samples) and UPLight~\cite{liu2025sharecmp} (1,991 samples). 
Both datasets introduce significant domain shift relative to the PIP dataset, particularly the underwater scenes in UPLight. 
Quantitative results are presented in \Cref{tab:results}, and qualitative examples appear in \Cref{fig:sim_res} (grayscale configuration). 
Our method consistently outperforms both physics-based and learning-based baselines across datasets, recovering richer high-frequency details and improving structural fidelity.

While the base model generalizes well, we further explore whether performance can be improved through fine-tuning on a small number of samples from the target domain. Specifically, for each of the ZJU-RGB-P and UPLight datasets, we fine-tune the PIP-trained model using 10 randomly selected pairs of RGB-polarization images for 5 epochs using the same training setup.
The qualitative results (\Cref{fig:sim_res}) show that an additional improvement in performance can be achieved using a small number of examples from the target domain. This makes domain adaptation more practical, without needing to collect a large dataset. The quantitative results are provided in the supplementary.

\subsection{Real-world Results}
\label{sec:results:real}

\begin{figure}[t]
\centering
\setlength{\tabcolsep}{0.8pt} %

\newcommand{\imgw}{0.195\columnwidth}

\begin{tabular}{*{5}{>{\centering\arraybackslash}m{\imgw}}}
{\small {RGB (Guide)}} &
{\small {FISTA Reconstruction}} &
{\small {FISTA + Transformer}} &
{\small {Ours}} &
{\small {Reference}} \\[3pt]

\includegraphics[width=\linewidth]{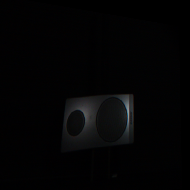} &
\includegraphics[width=\linewidth]{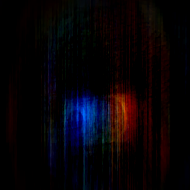} &
\includegraphics[width=\linewidth]{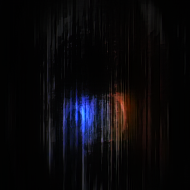} &
\includegraphics[width=\linewidth]{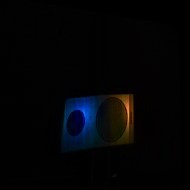} &
\includegraphics[width=\linewidth]{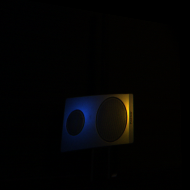} \\

\includegraphics[width=\linewidth]{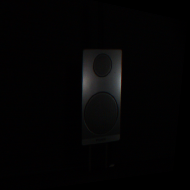} &
\includegraphics[width=\linewidth]{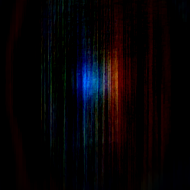} &
\includegraphics[width=\linewidth]{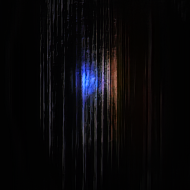} &
\includegraphics[width=\linewidth]{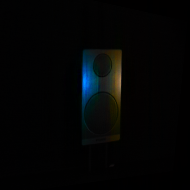} &
\includegraphics[width=\linewidth]{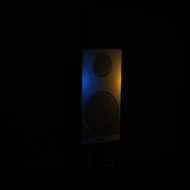} \\
\includegraphics[width=\linewidth]{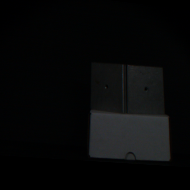} &
\includegraphics[width=\linewidth]{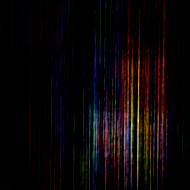} &
\includegraphics[width=\linewidth]{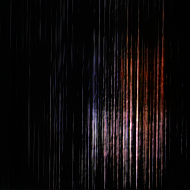} &
\includegraphics[width=\linewidth]{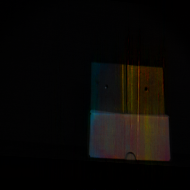} &
\includegraphics[width=\linewidth]{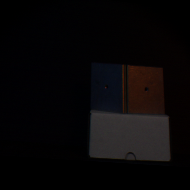} \\
\includegraphics[width=\linewidth]{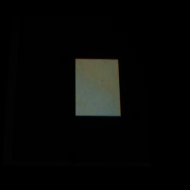} &
\includegraphics[width=\linewidth]{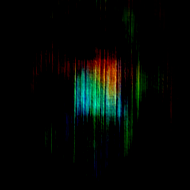} &
\includegraphics[width=\linewidth]{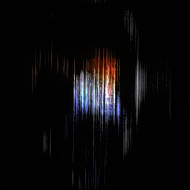} &
\includegraphics[width=\linewidth]{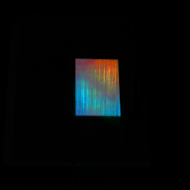} &
\includegraphics[width=\linewidth]{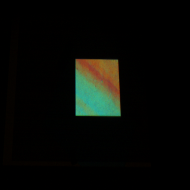} 
\end{tabular}
\vspace{-0.1in}
\caption{Qualitative results on real lensless polarization data (3-angle grayscale). Each reconstructed polarization triplet (\(0^\circ\), \(45^\circ\), \(90^\circ\)) is visualized as an RGB composite. Note the significant improvement in the structural details achieved by RGB guidance.}
\label{fig:realres}
\vspace{-0.2in}
\end{figure}

\begin{table*}[t]
\centering
\small
\caption{Ablation study results (PSNR $\uparrow$ / SSIM $\uparrow$ / LPIPS $\downarrow$). Additional details appear in the supplementary.}
\label{tab:fista_psfs}
\begin{tabular}{lccccccccc}
\hline
\textbf{Dataset \hspace{0.1cm} \textbackslash  \hspace{0.1cm} Model} & 
\multicolumn{3}{c}{\textbf{FISTA}} &
\multicolumn{3}{c}{\textbf{FISTA + Transf.}} &
\multicolumn{3}{c}{\textbf{Ours (FISTA input)}} \\
 & PSNR$\uparrow$ & SSIM$\uparrow$ & LPIPS$\downarrow$ & PSNR$\uparrow$ & SSIM$\uparrow$ & LPIPS$\downarrow$ & PSNR$\uparrow$ & SSIM$\uparrow$ & LPIPS$\downarrow$ \\
\hline
PIP (simple fusion) & 13.87 & 0.45 & 0.45 & - & - & - & 34.69 & 0.97 & 0.03 \\
UPLight (simple fusion) & 16.72 & 0.26 & 0.53 & - & - & - & 21.86 & 0.54 & 0.30 \\
ZJU-RGB-P (simple fusion) & 14.50 & 0.46 & 0.44 & - & - & - & 30.80 & 0.97 & 0.08 \\
UPLight (PSF \#1) & 15.01 & 0.27 & 0.62 & 17.96 & 0.34 & 0.63 & 20.78 & 0.53 & 0.32 \\
UPLight (PSF \#2) & 16.03 & 0.26 & 0.54 & 19.09 & 0.44 & 0.53 & 20.68 & 0.52 & 0.32 \\
ZJU-RGB-P (PSF \#1) & 14.45 & 0.45 & 0.47 & 24.28 & 0.83 & 0.27 & 31.01 & 0.97 & 0.08 \\
ZJU-RGB-P (PSF \#2) & 14.59 & 0.47 & 0.42 & 27.17 & 0.89 & 0.19 & 31.18 & 0.97 & 0.07 \\
PIP (5k iters eval, 10k model) & 13.09 & 0.43 & 0.48 & 23.61 & 0.83 & 0.17 & 34.83 & 0.97 & 0.03 \\
PIP  (5k iters eval, 5k model) & 13.09 & 0.43 & 0.48 & 27.53 & 0.87 & 0.14 & 34.93 & 0.97 & 0.03 \\
UPLight  (5k iters eval, 10k model) & 14.98 & 0.25 & 0.60 & 18.71 & 0.42 & 0.58 & 20.93 & 0.53 & 0.33 \\
UPLight  (5k iters eval, 5k model) & 14.98 & 0.25 & 0.60 & 16.32 & 0.39 & 0.56 & 21.17 & 0.53 & 0.32\\ 
UPLight (1k iters) & 11.88 & 0.20 & 0.72 & 14.99 & 0.36 & 0.71 & 20.91 & 0.53 & 0.43 \\
ZJU-RGB-P (5k iters eval, 10k model) & 14.45 & 0.46 & 0.45 & 25.77 & 0.87 & 0.21 & 31.12 & 0.97 & 0.08 \\
ZJU-RGB-P (5k iters eval, 5k model) & 14.45 & 0.46 & 0.45 & 26.01 & 0.87 & 0.21 & 30.92 & 0.97 & 0.08 \\
ZJU-RGB-P (1k iters) & 14.93 & 0.37 & 0.58 & 19.73 & 0.68 & 0.39 & 30.19 & 0.97 & 0.09 \\
\hline
\end{tabular}
\label{tab:ablations}
\vspace{-0.08in}
\end{table*}

We validate our method on data captured using the prototype lensless polarization camera, as shown qualitatively in \Cref{fig:realres}. 
We tested two polarized lighting setups: (i) front-illuminated scenes with two orthogonally polarized projectors (rows 1–3) to assess source separation, and (ii) a back-illuminated scene with a polarized screen and transparent plastic bag (row 4), highlighting polarization-dependent variations such as internal stress.
For the grayscale configuration, the polarization images are first reconstructed using FISTA with three angles provided as input to the refinement network
Despite the domain gap and differences in the polarization mask used for training, our RGB-guided method recovers sharper structures and finer details compared to FISTA. The FISTA+Transformer baseline fails to generalize in this setting, further highlighting the robustness and effectiveness of our proposed approach under real-world conditions. The prototype introduces additional challenges, including microfabrication artifacts in the polarization mask, the distance between the sensor and the polarization mask (due to the sensor's cover glass), and various fabrication and assembly inaccuracies that are not fully calibrated and compensated \cite{kraicer2026lensless}. These issues can lead to deviations from the assumed forward model, potentially degrading the performance of physics-based methods.

For real-world lensless measurements, pixel-wise aligned ground-truth polarization images are not available due to hardware differences between the reference and lensless cameras. The reference images in \Cref{fig:realres}, captured using a conventional RGB camera with a rotating polarizer, serve as qualitative benchmarks.
Unlike synthetic data, our real measurements require alignment. We perform a one-time homography-based alignment using manually selected correspondences, justified by the short inter-camera baseline relative to scene depth (Fig.~1, supplementary), assuming synchronized acquisition. However, residual misalignment may still persist. To handle this, we employ translation augmentation during training (see \Cref{subsec:network}). Additional qualitative results for the 4-angle RGB configuration are provided in the Supplementary Material.

\subsection{Ablation Studies}
\label{sec:results:ablation}
We analyze the impact of RGB fusion strategy, PSF mismatch, and FISTA depth under the three-angle grayscale setup on PIP (\Cref{tab:ablations}). Additional ablations (RGB-only and translation augmentation) are shown in the supplementary.

To assess the benefit of cross-attention for fusion versus a simpler alternative while retaining the advantages of the attention-based design, we evaluate a \textbf{simple fusion} variant. This variant replaces the cross-attention fusion with direct channel concatenation of the two modalities' features, followed by a $3\times3$ convolution and standard Swin Transformer blocks (RSTB).
Unlike our regular fusion module, which employs cross-attention (CRSTB) layers that jointly process both modalities through mutual attention before concatenation, this variant performs early feature-level merging without any adaptive interaction between modalities. Although quantitative differences are modest, the variant exhibits noticeable visual intensity artifacts on both simulated and real data (see qualitative results in Fig.~4, supplementary), similar to those reported in ~\cite{ma2022swinfusion}.

Next, we evaluate robustness to \textbf{PSF mismatch} by training on data simulated and reconstructed with our PSF, and testing on data simulated and reconstructed with two alternative PSFs. These PSFs are measured from different diffusers: the first, from \citet{antipa2018diffusercam}, exhibits a large rectangular pattern, while the second, from \citet{monakhova2020spectral}, covers a smaller rectangular area in comparison to our circular shape (see supplementary). \Cref{tab:fista_psfs} shows that our model consistently outperforms both FISTA and FISTA+Transformer under these mismatched conditions, demonstrating strong generalization to unseen optics.

We analyze sensitivity to the \textbf{number of FISTA iterations} used for the initial reconstruction at test time. Classical FISTA degrades rapidly when iterations are reduced (1k or 5k vs.\ the 10k  baseline used for both training and testing), reflecting its dependence on full convergence. In contrast, our method, leveraging RGB guidance, maintains high fidelity even with weak initializations, demonstrating robustness under limited test-time computation.
When trained with fewer iterations (5k instead of 10k), the method remains relatively stable under weaker initialization (\Cref{tab:ablations}). However, performance still drops compared to models trained and evaluated with fully converged (10k) inputs (\Cref{tab:results}). This highlights the importance of high-quality physics-based initialization and the trade-off between reconstruction accuracy and computational efficiency.

Finally, we evaluate the effect of \textbf{translation augmentation}. A model trained without it is sensitive to spatial shifts, whereas incorporating translations largely mitigates this effect, which is critical in real-world settings where perfect alignment cannot be guaranteed (Tab.~5, supplementary).

\section{Conclusions}
\label{sec:conclusions}
We introduced a modular two-stage framework for RGB-guided reconstruction in lensless polarization imaging, demonstrating high reconstruction quality, strong generalization, and robustness across both simulated and real-world setups. 
Our framework outperforms all compared baselines, confirming the contribution of our RGB-guided reconstruction approach. 
Beyond lensless setups, our approach can be extended to conventional polarization cameras with limited resolution or high noise, where RGB guidance can enhance the recovery of fine details. This modular design supports integration with alternative inverse solvers and adapts to a wide range of imaging setups. 
Future work will focus on improving computational efficiency and extending the framework toward dynamic scenes, advancing compact polarization imaging in practical environments.

\section*{Acknowledgment}
We thank Tomer Pee'r and Michael Baltaxe (General Motors) for providing a suitable version of the PIP dataset, and Shay Elmalem for fruitful discussions. This work was partially supported by the Center for AI and Data Science at Tel Aviv University (TAD) and by ERC Grant No.~10111339.

{
    \small
    \bibliographystyle{ieeenat_fullname}
    \bibliography{main}
}

\def\paperID{} %
\def\confName{CVPR}
\def\confYear{2026}

\title{Guided Lensless Polarization Imaging}

\maketitlesupplementary
\setcounter{figure}{0}
\setcounter{table}{0}
\setcounter{section}{0}
\setcounter{equation}{0}

This supplementary material provides additional details that supplement the paper \textit{``Guided Lensless Polarization Imaging''}. \Cref{sec:Appendixoptical} describes the optical setup used in our experiments. \Cref{sec:appendix_physics_based} outlines the implementation of the physics-based reconstruction baselines (FISTA and ADMM). \Cref{sec:appendix_SwinFuSR} presents additional details of the RGB-guided refinement network and \Cref{sec:appendix_results} summarizes further  implementation details and shows complementary results.

\begin{figure}[b]
\centering
\includegraphics[width=\linewidth]{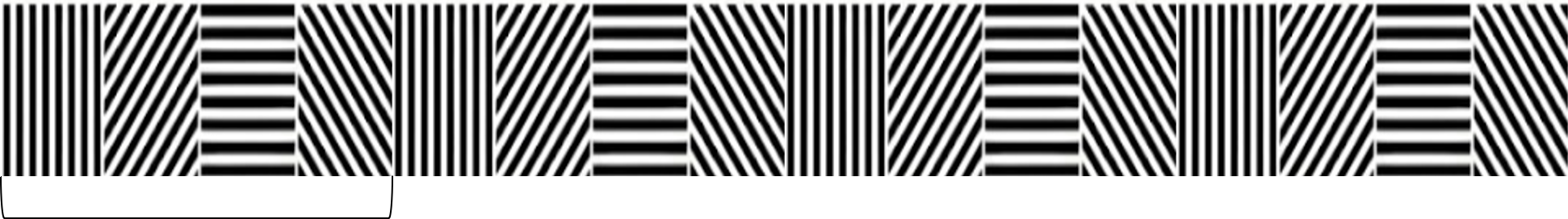}
\caption{Illustration of our polarization mask, containing four repetitions of striped polarizers at orientations $0^\circ, 45^\circ, 90^\circ,\text{and } 135^\circ$ }
\label{fig:mask}
\end{figure}

\section{Optical Setup}
\label{sec:Appendixoptical}
\begin{figure}[t]
\centering
\includegraphics[width=0.85\linewidth]{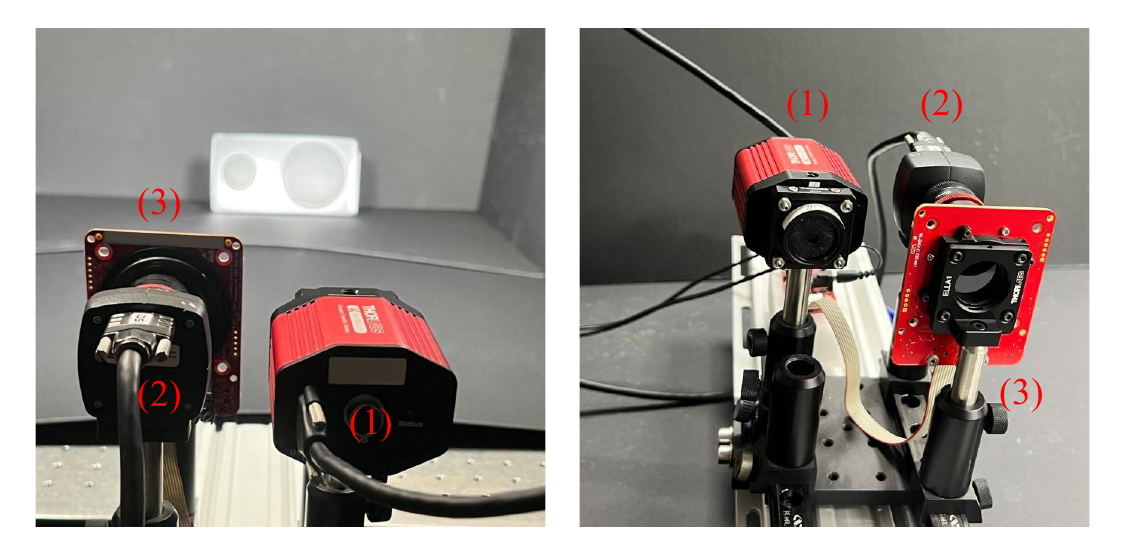}
\caption{The experimental optical setup from two viewpoints. \textbf{Left:} Front view showing the imaging target and the two-sensor setup: (1) lensless polarization camera and (2) RGB reference camera. \textbf{Right:} Back view of the setup showing (3) the rotation stage with a mounted linear polarizer positioned in front of (2) the RGB camera for capturing reference images. The lensless camera prototype (1) consists of a diffuser and a manually assembled polarization mask mounted on the sensor.}
\label{fig:opticsys}
\end{figure}

Our prototype lensless polarization camera, shown in \Cref{fig:opticsys}, combines a random diffuser with a polarization mask following the design of \citet{kraicer2026lensless,elmalem2021lensless}, as described in the Lensless Polarization Imaging Setup section of the main paper (Section 3.1).
These two optical elements jointly encode spatial and polarization information in the image plane. We use a $0.5^\circ$ diffuser (Edmund Optics \#47-860), mounted on a 12.3\,MP color CMOS sensor (Thorlabs CS126CU) with a pixel pitch of $3.45\,\mu\text{m}$.

The polarization mask is fabricated by cutting a linear polarizing film (Thorlabs LPVISE2X2) into stripes approximately \(880\,\mu\text{m}\) wide (corresponding to \(\sim\)256 sensor pixels). These stripes are placed on a glass substrate and arranged into the desired polarization-angle pattern, following a repeating sequence of \(0^\circ, 45^\circ, 90^\circ,\) and \(135^\circ\) (repeated four times), as illustrated in \Cref{fig:mask}. The mask is designed to support one-shot acquisition of the four polarization angles while remaining simple to fabricate under lab conditions, and is similar to prior designs \cite{kraicer2026lensless,elmalem2021lensless,baek2022lensless}. The mask assembly is then mounted just above the sensor’s cover glass to ensure accurate, high-fidelity polarization sampling at each angle.
Incoming light first passes through the diffuser and is multiplexed before reaching the polarization mask, after which the encoded light is recorded by the sensor.
To measure the system’s polarization-independent PSF, we image a point light source after removing the camera's polarization mask. To measure the angular response introduced by the polarization mask (at $0^\circ$, $45^\circ$, $90^\circ$, and $135^\circ$), we place a broadband, uniform light source in front of the bare lensless camera (after removing the diffuser) and control the incident polarization angle using an external motorized rotating linear polarizer (Thorlabs LPVISE100-A) mounted on a Thorlabs ELL14K stage. These angular response measurements are shown in Figure 3 of the main paper. We operate in a regime where the PSF is approximately shift-invariant (LSI), which simplifies our reconstruction to a linear shift-invariant model.
To obtain reference polarization images at each angle, an RGB sensor (UI-3590LE-C-HQ) is placed behind the rotating polarizer. These reference measurements are also used to compute the RGB guidance image, as defined in Eq.~(2) of the main paper, in the same manner as the simulated datasets. A photograph of the complete optical setup is shown in \Cref{fig:opticsys}.
The raw 16-bit lensless sensor data are converted to 32-bit floating point and normalized to the $[0, 1]$ range. White balance correction is then applied by scaling the red and blue channels to match the mean intensity of the green channel, before further processing.

\section{Physics-Based Reconstruction Implementation Details}
\label{sec:appendix_physics_based}
For both FISTA and ADMM, the polarization intensity image $\hat{\mathbf{x}}$ is recovered by solving the optimization problem in Eq.~(3) with its forward operator in Eq.~(1) of the main paper.
Both solvers are GPU-accelerated using CuPy.

\subsection{FISTA}
We adopt the FISTA solver with Haar-based anisotropic total variation (TV) from the publicly available implementation of SpectralDiffuserCam \cite{monakhova2020spectral}. 

For simulated data (PIP, UPLight, and ZJU-RGB-P), FISTA is run for 10k iterations, an empirically chosen value that ensures visual convergence and stable reconstruction quality.
For real-world measurements, the iteration count is reduced to 500 in both color and grayscale, which is sufficient for convergence. 

A fixed step size of $1/(L\cdot c)$ is used for FISTA's update, where $L$ is the Lipschitz constant of the forward operator and $c$ is a tuning factor. Following prior work, we set $c=45$ for synthetic data~\cite{monakhova2020spectral}, $c=100$ for the PSF-mismatch ablation, and use a more conservative $c=1000$ for real measurements to improve stability under noise and forward-model mismatch. All reconstructions are initialized with zeros.

We regularize $\mathbf{x}$ using a Haar-based anisotropic 3DTV prior across both spatial and polarization dimensions. Let $\lambda$ denote the global regularization strength and $\lambda_w$ the relative weighting of the polarization axis with respect to the spatial axes. The directional weights are defined as
\[
w_{\text{ax}} =
\begin{cases}
\lambda_w, & \text{if } \texttt{axis} = \text{polarization},\\
1,         & \text{if } \texttt{axis} = \text{spatial}.
\end{cases}
\]
Each FISTA iteration then performs a proximal update that combines non-negativity with Haar-based TV:
\[
\mathbf{x} \leftarrow \tfrac{1}{2} \Big(
\max(\mathbf{x}, 0) \;+\; 
\texttt{tvApproxHaar}\!\big(\mathbf{x}, \tfrac{\lambda}{L\cdot c}, w_{\text{ax}}\big)
\Big),
\]
Where the scaling factor $L\cdot c$ is applied consistently in both the gradient step and the TV thresholding.

We set $\lambda = \lambda_w = 5\times10^{-5}$ for synthetic data and increase both to $5\times10^{-3}$ for real measurements, which require stronger TV regularization. All hyperparameters were selected via grid search to balance denoising and structure preservation. For real data in the three-angle grayscale configuration, the measured PSF and mask are converted to grayscale before reconstruction.

\subsection{ADMM}
We also solve Eq.~(3) using scaled ADMM~\cite{boyd2011distributed}, with updates:
\begin{align}
(A^\top A + \rho I)\, v^{t+1} &= A^\top y + \rho\, (z^t - u^t), \label{eq:admm-v} \\[4pt]
z^{t+1} &= \operatorname{prox}_{(\lambda/\rho)\,\mathrm{TV}}\!\left(v^{t+1} + u^t\right), \label{eq:admm-z} \\[4pt]
u^{t+1} &= u^t + v^{t+1} - z^{t+1}. \label{eq:admm-u}
\end{align}

Here, $\mathbf{v}$ is the data-fidelity variable, $\mathbf{z}$ is the regularization variable enforcing the TV and non-negativity constraints, and $\mathbf{u}$ is the scaled dual variable (Lagrange multiplier) enforcing consensus between $\mathbf{v}$ and $\mathbf{z}$. The variables $\mathbf{A}$ and $\mathbf{y}$ represent the system's forward operator and the measurement vector, respectively, as defined in Eq.~(3) in the main paper.

Unlike FISTA, in ADMM the non-negativity constraint is applied to the TV-regularized variable $z$, i.e.,
$z^{t+1} \leftarrow \max\!\big(z^{t+1}, 0\big)$, rather than directly to $v$.

The TV proximal operator uses the same Haar-based anisotropic formulation as in the FISTA solver.
We solve \eqref{eq:admm-v} inexactly using conjugate gradients, since the forward operator  $\mathbf{A}$ includes spatial masking and cropping, making $A^\top A$
non-shift-invariant and thus not diagonalizable in the Fourier domain.
We use $\rho=0.15$, $\lambda=3\times10^{-5}$, $\lambda_w=6\times10^{-5}$, $200$ ADMM iterations, and CG tolerance $10^{-3}$ with $30$ inner iterations for all simulation datasets. We use ADMM only for simulated datasets in our ablations; all real-data reconstructions are obtained with FISTA.

\paragraph{Note on RGB Data.}
For both the FISTA and ADMM implementations, when processing four-angle RGB measurements, the 3D total-variation regularization (spatial and polarization dimensions) is applied separately to each color channel.

\section{RGB-Guided Deep Refinement Implementation Details}
\label{sec:appendix_SwinFuSR}
The second stage of our reconstruction pipeline utilizes an RGB-guided refinement network based on SwinFuSR \cite{arnold2024swinfusr}, as described in the RGB-Guided Deep Refinement section of the main paper (Section 3.4).

The network architecture follows the structure of SwinFuSR and is composed of three core modules: \textbf{Extraction}, \textbf{Fusion}, and \textbf{Reconstruction}. 

The Extraction module applies shallow convolutions and Swin Transformer Layers (STLs) to encode features from the initial reconstruction and RGB image.
The Fusion module integrates these feature streams through Attention-guided Cross-domain Fusion (ACF) blocks.
The Reconstruction module refines the fused representation with additional STLs and convolutions to produce the final output with a skip connection for the initial reconstruction.

The module depths are configured as follows: two STLs in the Extraction module, three ACF blocks in the Fusion module, and three STLs in the Reconstruction module, consistent with the original SwinFuSR configuration.

The original implementation of SwinFuSR was designed for guided thermal Super-Resolution (SR), where the low-resolution (LR) thermal input image is upsampled to high-resolution (HR) before entering the network. We omit this upsampling step because both our initial reconstruction and the RGB guidance image are already at the target high resolution, making the operation unnecessary.
We used a batch size of 1 due to memory constraints.

\section{Additional Experimental Results}
\label{sec:appendix_results}
This section details complementary information for the experimental results reported in Section 4 of the main paper.

\subsection{Additional Implementation Details}

\paragraph{Training Dataset}
We selected the Polarimetric Imaging for Perception (PIP) dataset \cite{baltaxe2023polarimetric} because it is the largest publicly available dataset providing aligned RGB and polarization data suitable for our task. However, the publicly released version contains only derived polarization metrics (Angle of Linear Polarization (AoLP) and Degree of Linear Polarization (DoLP)) along with RGB images, which are computed from the raw polarization intensity images ($I_0, I_{45}, I_{90}, I_{135}$). Since our method requires these raw intensity measurements, we obtained them directly from the dataset authors upon request, along with their established preprocessing pipeline.
We split the data into 8,538 training, 2,717 validation, and 1,372 test images without scene overlap.

\paragraph{Evaluation Metrics}
We used three metrics for quantitative evaluation: PSNR, SSIM, and LPIPS (VGG). For the three-angle grayscale polarization intensity images configuration, PSNR and SSIM are averaged over channels, while LPIPS is computed per channel (converted to RGB) and averaged.
For the 12-channel RGB four-angle configuration, PSNR and SSIM are averaged over all 12 channels. LPIPS is computed per RGB triplet (i.e., three channels representing a single polarization angle) and then averaged over the four triplets.
 
\begin{table}[t!]
\centering
\setlength{\tabcolsep}{4pt}
\caption{
Baseline comparison under the three-angle grayscale configuration on the UPLight and ZJU-RGB-P evaluation datasets.
}
\label{tab:baseline_two_sets}
\begin{tabular}{lccc}
\toprule
Model & PSNR$\uparrow$ & SSIM$\uparrow$ & LPIPS$\downarrow$ \\
\midrule
\multicolumn{4}{c}{\textbf{UPLight}} \\
FlatNet & 10.78 & 0.27 & 0.98 \\
PolarAnything (FISTA input) & 11.84 & 0.36 & 0.98 \\
PolarAnything (RGB input) & 11.98 & 0.40 & 0.93 \\
FISTA & 16.72 & 0.26 & 0.53 \\
FISTA + Transf. & 17.93 & 0.44 & 0.53 \\
Ours (FISTA input) & \textbf{20.49} & \textbf{0.52} & \textbf{0.32} \\
\midrule
\multicolumn{4}{c}{\textbf{ZJU-RGB-P}} \\
FlatNet & 16.73 & 0.54 & 0.57 \\
PolarAnything (FISTA input) & 19.05 & 0.58 & 0.42 \\
PolarAnything (RGB input) & 19.96 & 0.62 & 0.38 \\
FISTA  & 14.50 & 0.46 & 0.44 \\
FISTA + Transf. & 27.20 & 0.89 & 0.19 \\
Ours (FISTA input) & \textbf{31.19} & \textbf{0.97} & \textbf{0.07} \\
\bottomrule
\end{tabular}
\end{table}

\subsection{Synthetic Data Results}
This section further elaborates on the synthetic data results presented in Section 4.1 of the main paper.

\paragraph{FlatNet and PolarAnything}
FlatNet \cite{khan2020flatnet} and PolarAnything \cite{zhang2025polaranything} are two additional baselines included in our comparisons, as detailed in the Experimental Results section of the main paper (Section 4.1).

FlatNet provides both separable and non-separable variants depending on the structure of the PSF. We use the non-separable model, which aligns with the characteristics of our measured PSF. Because its U-Net architecture expects input dimensions divisible by 32, we pad the 250×250 sensor images to the nearest valid size (256×256). 
Training was performed for 50 epochs (approximately the same number of iterations in the paper). The rest of the parameters were the same as their code's base config and only MSE loss, which yielded the best results.

For PolarAnything, we likewise pad the 250×250 inputs to 256×256 to satisfy the spatial-resolution requirements of the diffusion U-Net. We train the network for each configuration (RGB / FISTA input) for 20 epochs, which is sufficient for convergence; all other training parameters follow those reported in the paper. PolarAnything was trained on two NVIDIA RTX A6000 GPUs with a batch size of 32.

For both baselines, the reconstructed images are cropped to their original resolution before computing all evaluation metrics.
\Cref{tab:baseline_two_sets} presents the generalization performance of FlatNet and PolarAnything on the supplementary UPLight and ZJU-RGB-P datasets, which were not featured in the main paper (see Table 2 in the main paper). 
Both models show limited generalization, notably on the UPLight dataset, where their results are worse than the FISTA baseline, clearly demonstrating a significant performance gap compared to our proposed approach and the unguided FISTA+Transformer baseline.

\setlength{\tabcolsep}{0.5pt}
\renewcommand{\arraystretch}{0.99}
\newcolumntype{M}[1]{>{\centering\arraybackslash}m{#1}}
\newcolumntype{C}[1]{>{\centering\arraybackslash}m{#1}}
\newlength\polH
\setlength{\polH}{3.4cm}
\newcommand{\polimg}[1]{\includegraphics[width=\linewidth,height=\polH,keepaspectratio]{#1}}
\newcommand{\rgbimg}[1]{\includegraphics[width=\linewidth,height=\polH,keepaspectratio]{#1}}

\begin{figure*}[t]
\centering

\begin{subfigure}[t]{0.48\textwidth}
\centering\scriptsize
\begin{tabular}{@{} M{0.06\linewidth}
  C{0.18\linewidth} C{0.18\linewidth} C{0.18\linewidth}
  C{0.18\linewidth} C{0.18\linewidth} @{}} 
 & \textbf{FISTA} & \makecell{\textbf{FISTA}\\ \textbf{+ Transformer}}
 & \textbf{Ours} & \textbf{Reference} & \textbf{RGB} \\

$0^\circ$ &
\polimg{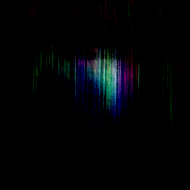} &
\polimg{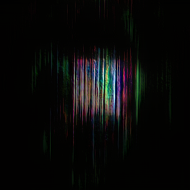} &
\polimg{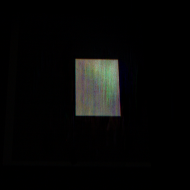} &
\polimg{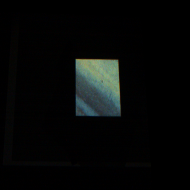} &
\multirow{4}{*}[-0.5\polH]{\rgbimg{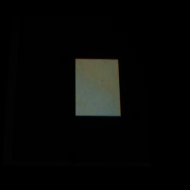}} \\

$45^\circ$ &
\polimg{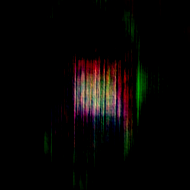} &
\polimg{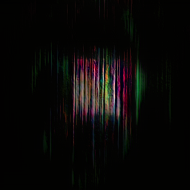} &
\polimg{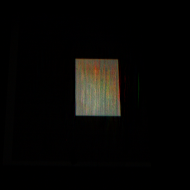} &
\polimg{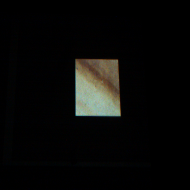} \\

$90^\circ$ &
\polimg{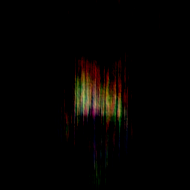} &
\polimg{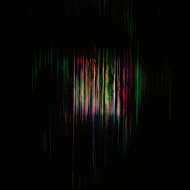} &
\polimg{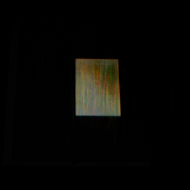} &
\polimg{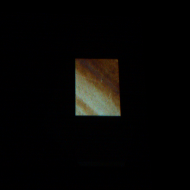} \\

$135^\circ$ &
\polimg{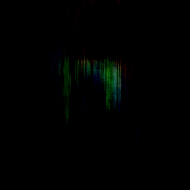} &
\polimg{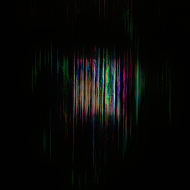} &
\polimg{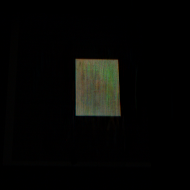} &
\polimg{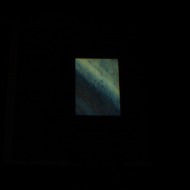} \\
\end{tabular}
\caption{Plastic bag}
\end{subfigure}
\hspace{0.015\textwidth}
\begin{subfigure}[t]{0.48\textwidth}
\centering\scriptsize
\begin{tabular}{@{} M{0.06\linewidth}
  C{0.18\linewidth} C{0.18\linewidth} C{0.18\linewidth}
  C{0.18\linewidth} C{0.18\linewidth} @{}} 
 & \textbf{FISTA} & \makecell{\textbf{FISTA}\\ \textbf{+ Transformer}}
 & \textbf{Ours} & \textbf{Reference} & \textbf{RGB} \\

$0^\circ$ &
\polimg{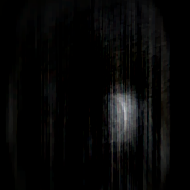} &
\polimg{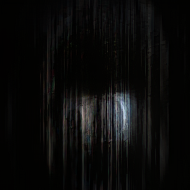} &
\polimg{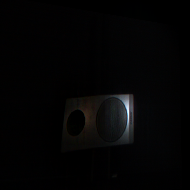} &
\polimg{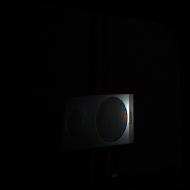} &
\multirow{4}{*}[-0.5\polH]{\rgbimg{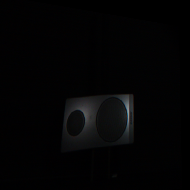}} \\

$45^\circ$ &
\polimg{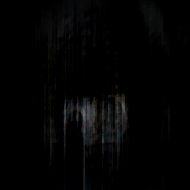} &
\polimg{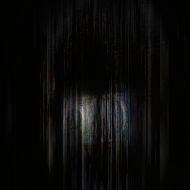} &
\polimg{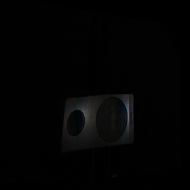} &
\polimg{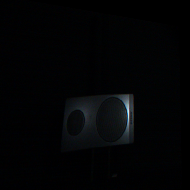} \\

$90^\circ$ &
\polimg{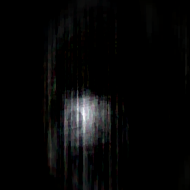} &
\polimg{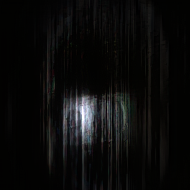} &
\polimg{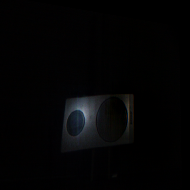} &
\polimg{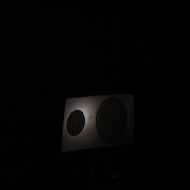} \\

$135^\circ$ &
\polimg{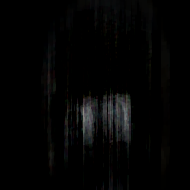} &
\polimg{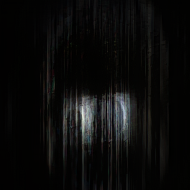} &
\polimg{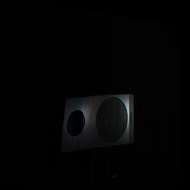} &
\polimg{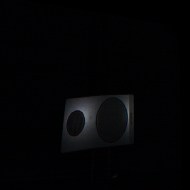} \\
\end{tabular}
\caption{Speaker (horizontal)}
\end{subfigure}

\vspace{0.2cm}

\begin{subfigure}[t]{0.48\textwidth}
\centering\scriptsize
\begin{tabular}{@{} M{0.06\linewidth}
  C{0.18\linewidth} C{0.18\linewidth} C{0.18\linewidth}
  C{0.18\linewidth} C{0.18\linewidth} @{}} 
 & \textbf{FISTA} & \makecell{\textbf{FISTA}\\ \textbf{+ Transformer}}
 & \textbf{Ours} & \textbf{Reference} & \textbf{RGB} \\

$0^\circ$ &
\polimg{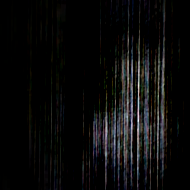} &
\polimg{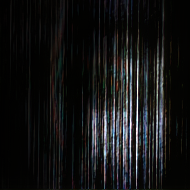} &
\polimg{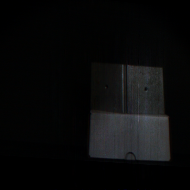} &
\polimg{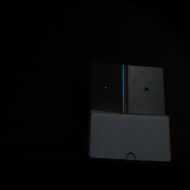} &
\multirow{4}{*}[-0.5\polH]{\rgbimg{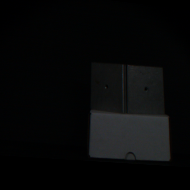}} \\

$45^\circ$ &
\polimg{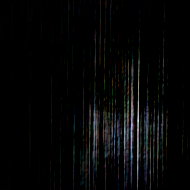} &
\polimg{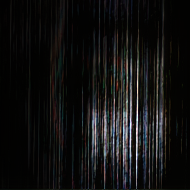} &
\polimg{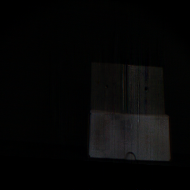} &
\polimg{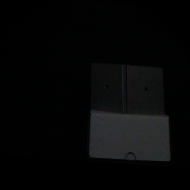} \\

$90^\circ$ &
\polimg{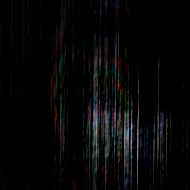} &
\polimg{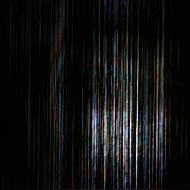} &
\polimg{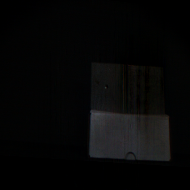} &
\polimg{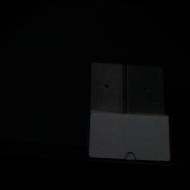} \\

$135^\circ$ &
\polimg{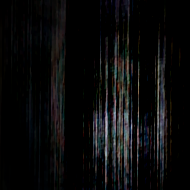} &
\polimg{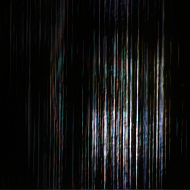} &
\polimg{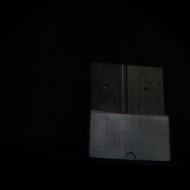} &
\polimg{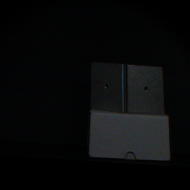} \\
\end{tabular}
\caption{Metals}
\end{subfigure}
\hspace{0.015\textwidth}
\begin{subfigure}[t]{0.48\textwidth}
\centering\scriptsize
\begin{tabular}{@{} M{0.06\linewidth}
  C{0.18\linewidth} C{0.18\linewidth} C{0.18\linewidth}
  C{0.18\linewidth} C{0.18\linewidth} @{}} 
 & \textbf{FISTA} & \makecell{\textbf{FISTA}\\ \textbf{+ Transformer}}
 & \textbf{Ours} & \textbf{Reference} & \textbf{RGB} \\

$0^\circ$ &
\polimg{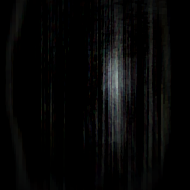} &
\polimg{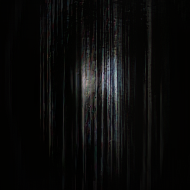} &
\polimg{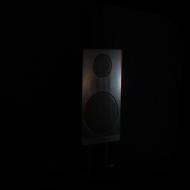} &
\polimg{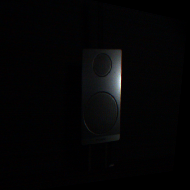} &
\multirow{4}{*}[-0.5\polH]{\rgbimg{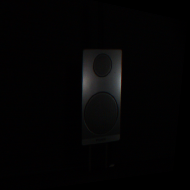}} \\

$45^\circ$ &
\polimg{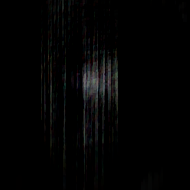} &
\polimg{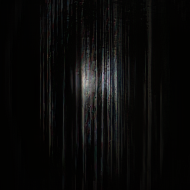} &
\polimg{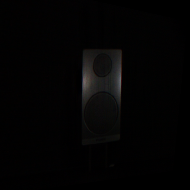} &
\polimg{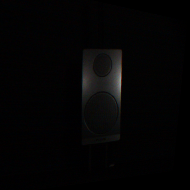} \\

$90^\circ$ &
\polimg{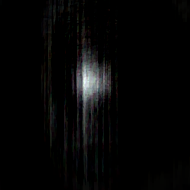} &
\polimg{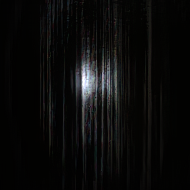} &
\polimg{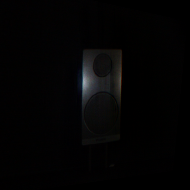} &
\polimg{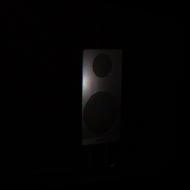} \\

$135^\circ$ &
\polimg{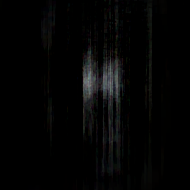} &
\polimg{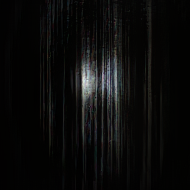} &
\polimg{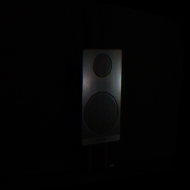} &
\polimg{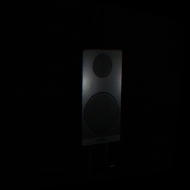} \\
\end{tabular}
\caption{Speaker (vertical)}
\end{subfigure}
\caption{Qualitative comparison on four real scenes: plastic bag, horizontal speaker, metals, and vertical speaker. For each scene, rows correspond to the four polarization angles, and columns show FISTA, FISTA+Transformer, our method, the reference polarization image, and the RGB guidance image used by our method (rightmost column).}
\label{fig:col_results_all}
\end{figure*}

\paragraph{Fine-Tuning Evaluation.}
\Cref{tab:finetune_res} shows the performance on \textbf{ZJU-RGB-P and UPLight} before and after fine-tuning on 10 image pairs, demonstrating quantitative gains with minimal target-domain data.
\Cref{tab:finetune_res_pip} summarizes the performance on \textbf{PIP} before and after this fine-tuning.  
As expected, domain shifts lead to degradation on the source domain (PIP), more pronounced for UPLight due to its larger domain shift (underwater scenes vs street scenes in PIP and ZJU-RGB-P). However, this degradation is moderate and represents a trade-off that enables improved performance on the target domain using only a small amount of new data.
\subsection{Real-world Results}
The qualitative results on real lensless polarization data in the 3-angle grayscale configuration are shown in Figure 5 in the main paper. For the same scenes, the results under the four-angle RGB configuration further confirm that our method consistently surpasses the reference approaches, yielding accurate and reliable reconstructions as seen in \Cref{fig:col_results_all}.

\begin{table}[t]
\centering
\setlength{\tabcolsep}{4pt}
\caption{Performance before and after fine-tuning on 10 training pairs from ZJU-RGB-P or UPLight, under the \textbf{four-angle RGB} and \textbf{three-angle grayscale} configurations, using the base model with FISTA input.}
\begin{tabular}{lccc ccc}
\toprule
\multirow{2}{*}{\textbf{Model}} &
\multicolumn{3}{c}{\textbf{UPLight}} &
\multicolumn{3}{c}{\textbf{ZJU-RGB-P}} \\
 & \footnotesize PSNR$\uparrow$ & \footnotesize SSIM$\uparrow$ & \footnotesize LPIPS$\downarrow$ & \footnotesize PSNR$\uparrow$ & \footnotesize SSIM$\uparrow$ & \footnotesize LPIPS$\downarrow$ \\
\midrule
\multicolumn{7}{c}{\textbf{Color}} \\
Base       & 20.06 & 0.51 & 0.28 & 30.36 & 0.96 & 0.04 \\
Fine-tuned & \textbf{21.99} & \textbf{0.55} & \textbf{0.18} & \textbf{31.66} & \textbf{0.97} & \textbf{0.03} \\
\midrule
\multicolumn{7}{c}{\textbf{Grayscale}} \\
Base       & 20.49 & 0.52 & 0.32 & 31.16 & \textbf{0.97} & 0.07 \\
Fine-tuned & \textbf{24.67} & \textbf{0.56} & \textbf{0.23} & \textbf{32.74} & \textbf{0.97} & \textbf{0.05} \\
\bottomrule
\end{tabular}
\label{tab:finetune_res}
\end{table}

\begin{table}[t]
\centering
\setlength{\tabcolsep}{4pt}
\caption{\textbf{Performance on the PIP dataset} before and after fine-tuning on 10 training pairs from UPLight or ZJU-RGB-P, evaluated under the four-angle RGB and three-angle grayscale configurations using the base model with FISTA input.}
\begin{tabular}{lccc}
\toprule
Model & \footnotesize PSNR$\uparrow$ & \footnotesize SSIM$\uparrow$ & \footnotesize LPIPS$\downarrow$ \\
\midrule
\multicolumn{4}{c}{\textbf{Color} }\\
Base - before                    & 33.05 & 0.95 & 0.04 \\
FT (UPLight)    & 28.95 & 0.92 & 0.06 \\
FT (ZJU-RGB-P)  & 31.53 & 0.95 & 0.05 \\
\midrule
\multicolumn{4}{c}{\textbf{Grayscale}} \\
Base - before                       & 35.13 & 0.97 & 0.03 \\
FT (UPLight)    & 31.26 & 0.96 & 0.07 \\
FT (ZJU-RGB-P)  & 33.53 & 0.97 & 0.04 \\
\bottomrule
\end{tabular}
\label{tab:finetune_res_pip}
\end{table}

\begin{figure*}[t!]
\centering
\renewcommand{\arraystretch}{0.8}
\setlength{\tabcolsep}{1pt}

\begin{minipage}[t]{0.49\textwidth}
\centering
\begin{tabular}{@{}m{0.10\linewidth} m{0.26\linewidth} m{0.26\linewidth} m{0.26\linewidth}@{}}

\multicolumn{1}{c}{} &
\makebox[0.88\linewidth][c]{\textbf{Simple fusion}} &
\makebox[0.88\linewidth][c]{\textbf{Ours}} &
\makebox[0.88\linewidth][c]{\textbf{GT}} \\[4pt]

\multirow{2}{*}{\rotatebox[origin=c]{90}{\textbf{UPLight}}} &
\includegraphics[width=0.85\linewidth]{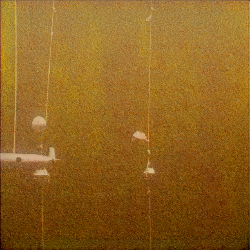} &
\includegraphics[width=0.85\linewidth]{imgs/image_1001_sim_pred_wmisalign_training.png} &
\includegraphics[width=0.85\linewidth]{imgs/image_1001_gt_pol.png} \\[4pt]

&
\includegraphics[width=0.85\linewidth]{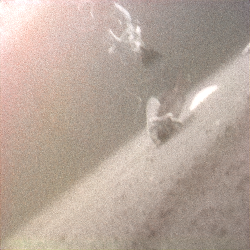} &
\includegraphics[width=0.85\linewidth]{imgs/image_1907_sim_pred_wmisalign_training.png} &
\includegraphics[width=0.85\linewidth]{imgs/image_1907_gt_pol.png} \\[8pt]

\multirow{2}{*}{\rotatebox[origin=c]{90}{\textbf{ZJU-RGB-P}}} &
\includegraphics[width=0.85\linewidth]{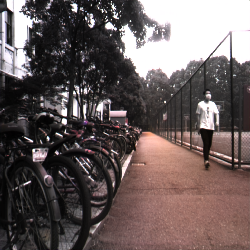}&
\includegraphics[width=0.85\linewidth]{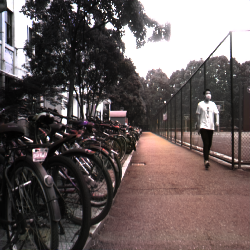} &
\includegraphics[width=0.85\linewidth]{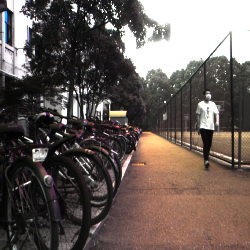} \\[4pt]

&
\includegraphics[width=0.85\linewidth]{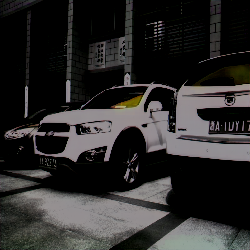} &
\includegraphics[width=0.85\linewidth]{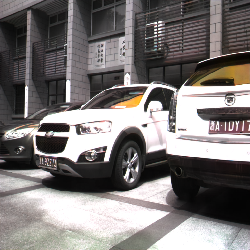} &
\includegraphics[width=0.85\linewidth]{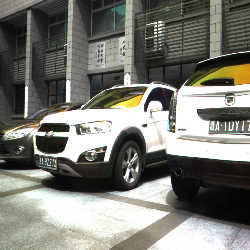}

\end{tabular}
\end{minipage}
\hfill
\begin{minipage}[t]{0.49\textwidth}
\centering
\begin{tabular}{@{}m{0.10\linewidth} m{0.26\linewidth} m{0.26\linewidth} m{0.26\linewidth}@{}}

\multicolumn{1}{c}{} &
\makebox[0.88\linewidth][c]{\textbf{Simple fusion}} &
\makebox[0.88\linewidth][c]{\textbf{Ours}} &
\makebox[0.88\linewidth][c]{\textbf{GT}} \\[4pt]

&
\includegraphics[width=0.85\linewidth]{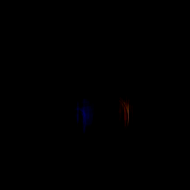} &
\includegraphics[width=0.85\linewidth]{imgs/real_imgs_f/ramkolh_gray_new_pred_wmisalignmenttraining.png} &
\includegraphics[width=0.85\linewidth]{imgs/real_imgs_f/RAMKOLHgraypolgtaligned.png} \\[4pt]

&
\includegraphics[width=0.85\linewidth]{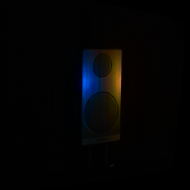} &
\includegraphics[width=0.85\linewidth]{imgs/real_imgs_f/ramkolv_gray_new_pred_wmisalignmenttraining.png} &
\includegraphics[width=0.85\linewidth]{imgs/RAMKOLVgraypolgtaligned.png} \\[4pt]

&
\includegraphics[width=0.85\linewidth]{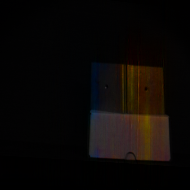} &
\includegraphics[width=0.85\linewidth]{imgs/real_imgs_f/metals_gray_new_pred_wmisalignmenttraining.png} &
\includegraphics[width=0.85\linewidth]{imgs/metal_gt_gray_aligned.png} \\[4pt]

&
\includegraphics[width=0.85\linewidth]{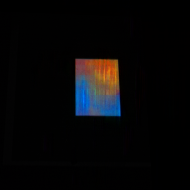} &
\includegraphics[width=0.85\linewidth]{imgs/real_imgs_f/PLASIC_gray_new_pred_wmisalignmenttraining.png} &
\includegraphics[width=0.85\linewidth]{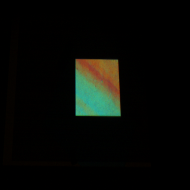}

\end{tabular}
\end{minipage}

\caption{
Qualitative comparison of \textbf{simple fusion} and \textbf{our method}.  
Left: simulated datasets (UPLight, ZJU-RGB-P).  
Right: real scenes.  
Each polarization triplet ($0^\circ$, $45^\circ$, $90^\circ$) is visualized as an RGB composite.
}
\label{fig:simple_fusion_all}
\end{figure*}

\subsection{Ablation Studies} 
This section provides supplementary details for the ablation studies in Section 4.3 of the main paper.
\paragraph{Simple Fusion}
While Table 4 in the main paper shows that simple feature fusion does not result in a significant quantitative performance drop compared to our method (which uses cross-attention fusion), the qualitative differences are present. As illustrated in \Cref{fig:simple_fusion_all}, simple fusion yields results with visual intensity that appears different and less accurate than our approach across both the simulated (UPLight, ZJU-RGB-P) and real-scene datasets. Furthermore, in the real-scene example of the first row, this simple fusion mechanism fails to effectively integrate the RGB and initial reconstruction features.

\begin{figure}[h!]
\centering
\includegraphics[width=0.9\linewidth]{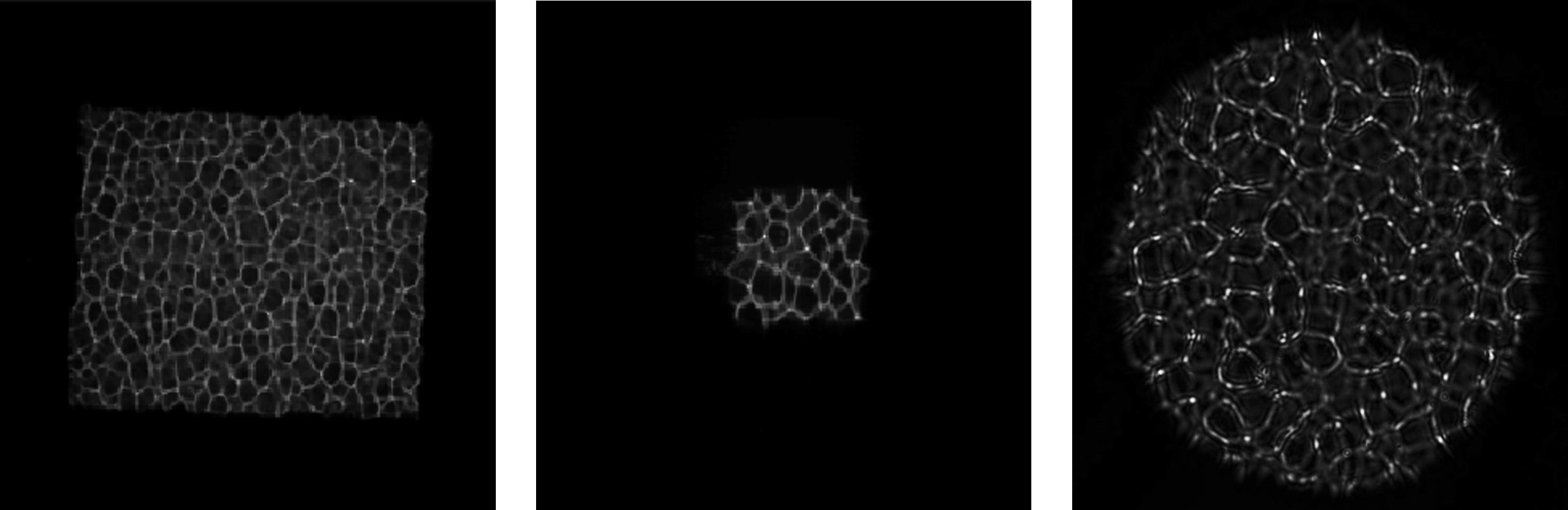}
\caption{PSFs used in ablation study (\#1 and \#2) and the training PSF (measured from our system). \textbf{Left:} PSF \#1. \textbf{Middle:} PSF \#2. \textbf{Right:} Measured PSF used for training.}
\label{fig:ablations_psfs}
\end{figure}

\paragraph{PSF Mismatch}
The results presented in Table 4 of the main paper demonstrate our model's robust generalization to unseen optics during inference. Specifically, we used two additional Point Spread Functions (PSFs) from \citet{antipa2018diffusercam} (PSF \#1) and \citet{monakhova2020spectral}  (PSF \#2) to simulate and reconstruct the UPLight and ZJU-RGB-P datasets. These PSFs are distinct from the training simulation and reconstruction PSF used for the PIP dataset. The visual comparison of all three PSFs is provided in \Cref{fig:ablations_psfs}. All PSFS share a similar speckle-like structure but differ in geometry.

\paragraph{RGB Guidance}
Similar to the FISTA+Transformer baseline, we train an RGB+Transformer model to highlight the limitations of RGB-only reconstruction.
The RGB-only model (\Cref{tab:rgb_ablation}) underperforms our method and fails to generalize to UPLight (an out-of-distribution (OOD) dataset), as it lacks polarization information. 
In contrast, our method preserves consistency with the physics-based polarization initialization via a skip connection, while RGB provides complementary high-frequency structure through cross-attention (see Fig.~4 in the main paper).

\paragraph{Translation Augmentation}
We evaluate robustness to misalignment by applying random translations to the RGB guidance image on simulated data (\Cref{tab:translation_robustness}). Without augmentation, performance degrades under such shifts, whereas training with translation augmentation largely eliminates this sensitivity. 

We further evaluate a wider range of translation magnitudes on simulated data and observe stable performance even beyond $\pm 4$ pixels. However, in real-world experiments, we find that using $\pm 4$ pixels during training provides the best overall performance, reflecting the typical level of residual misalignment in our setup. This is particularly important in practice, where perfect alignment cannot be guaranteed, motivating its use in the main paper.

\begin{table}[t]
\centering
\setlength{\tabcolsep}{5pt}
\renewcommand{\arraystretch}{1.05}
\caption{RGB ablation under the same training pipeline as our method. FISTA+T denotes the baseline derived from our architecture, in which the Transformer refines only the FISTA reconstruction without RGB guidance (i.e., the same input is fed to both branches). RGB+T uses the same architecture with RGB-only input. UPLight is out-of-distribution (OOD) relative to the training data.}
\label{tab:rgb_ablation}
\begin{tabular}{lcccc}
\toprule
\textbf{Model} &
\multicolumn{2}{c}{\textbf{PIP}} &
\multicolumn{2}{c}{\textbf{UPLight}} \\
 & PSNR$\uparrow$ & SSIM$\uparrow$ & PSNR$\uparrow$ & SSIM$\uparrow$ \\
\midrule
Polar (FISTA+T) & 28.85 & 0.88 & 17.93 & 0.44 \\
RGB (RGB+T)     & 32.28 & \textbf{0.97} & 13.11 & 0.48 \\
\textbf{Ours}   & \textbf{35.13} & \textbf{0.97} & \textbf{20.49} & \textbf{0.52} \\
\bottomrule
\end{tabular}
\end{table}

\newcommand{\pslash}{\,/\,}
\begin{table}[h]
\centering
\caption{Robustness to test-time translations (total shift: 0--4 px). Metrics are reported as PSNR / SSIM, averaged over two seeds.}
\label{tab:translation_robustness}
\small
\setlength{\tabcolsep}{3pt}
\renewcommand{\arraystretch}{0.88}
\begin{tabular}{l|ccccc}
\hline
\textbf{Dataset} & \textbf{0px} & \textbf{1px} & \textbf{2px} & \textbf{3px} & \textbf{4px} \\
\hline
\multicolumn{6}{l}{\emph{\textbf{Ours (no translation augmentation, grayscale)}}} \\
UPLight  & 19.92\pslash.54 & 19.84\pslash.52 & 19.74\pslash.51 & 19.61\pslash.50 & 19.46\pslash.50 \\
RGBP & 31.74\pslash.97 & 25.84\pslash.94 & 23.65\pslash.91 & 22.56\pslash.90 & 21.94\pslash.89 \\
PIP  & 35.53\pslash.97 & 27.91\pslash.94 & 25.36\pslash.92 & 24.05\pslash.91 & 23.23\pslash.90 \\
\hline
\multicolumn{6}{l}{\emph{\textbf{Ours (trained with translation augmentation, grayscale)}}} \\
UPLight  & 20.49\pslash.52 & 20.49\pslash.52 & 20.48\pslash.52 & 20.47\pslash.52 & 20.46\pslash.52 \\
RGBP & 31.19\pslash.97 & 31.15\pslash.97 & 31.10\pslash.97 & 31.03\pslash.97 & 30.95\pslash.97 \\
PIP  & 35.13\pslash.97 & 35.08\pslash.97 & 35.03\pslash.97 & 34.96\pslash.97 & 34.88\pslash.97 \\
\hline
\end{tabular}
\end{table}

\end{document}